\documentclass[11pt,journal=jacsat,manuscript=article,layout=twocolumn]{achemso}

\usepackage{chemformula} 
\usepackage[T1]{fontenc}
\usepackage{amsmath}
\usepackage{amssymb}
\usepackage{graphicx}
\usepackage{wasysym}
\usepackage[unicode=true,
 bookmarks=true,bookmarksnumbered=false,bookmarksopen=false,
 breaklinks=false,pdfborder={0 0 1},backref=false,colorlinks=false,pdfpagemode=FullScreen]{hyperref}
\hypersetup{pdfauthor={Clementi}}

\author{Jiang Wang}
\affiliation[CTBP]{Rice University, Center for Theoretical Biological Physics, Houston,
Texas 77005, United States}
\alsoaffiliation[RiceChemistry]{Rice University, Department of Chemistry, Houston, Texas 77005, United
States}

\author{Simon Olsson}
\affiliation[FU]{Freie Universit\"at Berlin, Department of Mathematics and Computer
Science, Arnimallee 6, 14195 Berlin, Germany}

\author{Christoph Wehmeyer}
\affiliation[FU]{Freie Universit\"at Berlin, Department of Mathematics and Computer
Science, Arnimallee 6, 14195 Berlin, Germany}

\author{Adri\`a P\'erez}
\affiliation[UPF]{Computational Science Laboratory, Universitat Pompeu Fabra, PRBB, C/
Dr Aiguader 88, 08003, Barcelona, Spain}

\author{Nicholas E. Charron}
\affiliation[CTBP]{Rice University, Center for Theoretical Biological Physics, Houston,
Texas 77005, United States}
\alsoaffiliation[RicePhysics]{Rice University, Department of Physics, Houston, Texas 77005, United
States}

\author{Gianni de Fabritiis}
\affiliation[UPF]{Computational Science Laboratory, Universitat Pompeu Fabra, PRBB, C/
Dr Aiguader 88, 08003, Barcelona, Spain}
\alsoaffiliation[ICREA]{Institucio Catalana de Recerca i Estudis Avanats (ICREA), Passeig
Lluis Companys 23, Barcelona 08010, Spain}

\author{Frank No\'e}
\affiliation[FU]{Freie Universit\"at Berlin, Department of Mathematics and Computer
Science, Arnimallee 6, 14195 Berlin, Germany}
\alsoaffiliation[CTBP]{Rice University, Center for Theoretical Biological Physics, Houston,
Texas 77005, United States}
\alsoaffiliation[RiceChemistry]{Rice University, Department of Chemistry, Houston, Texas 77005, United
States}
\email{frank.noe@fu-berlin.de}

\author{Cecilia Clementi}
\affiliation[CTBP]{Rice University, Center for Theoretical Biological Physics, Houston,
Texas 77005, United States}
\alsoaffiliation[RiceChemistry]{Rice University, Department of Chemistry, Houston, Texas 77005, United
States}
\alsoaffiliation[RicePhysics]{Rice University, Department of Physics, Houston, Texas 77005, United
States}
\alsoaffiliation[FU]{Freie Universit\"at Berlin, Department of Mathematics and Computer
Science, Arnimallee 6, 14195 Berlin, Germany}
\email{cecilia@rice.edu}

\title{Machine Learning of coarse-grained Molecular Dynamics Force Fields}

\begin{document}

\begin{abstract}
Atomistic or ab-initio molecular dynamics simulations are widely used
to predict thermodynamics and kinetics and relate them to molecular
structure. A common approach to go beyond the time- and length-scales
accessible with such computationally expensive simulations is the
definition of coarse-grained molecular models. Existing coarse-graining
approaches define an effective interaction potential to match defined
properties of high-resolution models or experimental data. In this
paper, we reformulate coarse-graining as a supervised machine learning
problem. We use statistical learning theory to decompose the coarse-graining
error and cross-validation to select and compare the performance of
different models. We introduce CGnets, a deep learning approach, that
learns coarse-grained free energy functions and can be trained by
a force matching scheme. CGnets maintain all physically relevant
invariances and allow one to incorporate prior physics knowledge to avoid
sampling of unphysical structures. 
We show that CGnets can capture
all-atom explicit-solvent free energy surfaces with models using only
a few coarse-grained beads and no solvent, while classical coarse-graining
methods fail to capture crucial features of the free energy surface.
Thus, CGnets are able to capture multi-body terms that emerge
from the dimensionality reduction.
\end{abstract}
\maketitle

\section{Introduction}

Recent technological and methodological advances have made possible
to simulate macromolecular systems on biologically relevant timescales
\citep{LindorffLarsenEtAl_Science11_AntonFolding,BuchEtAl_JCIM10_GPUgrid,ShirtsPande_Science2000_FoldingAtHome}.
For instance, one can simulate binding, folding and conformation changes
of small to intermediate-size proteins on timescales of milliseconds,
seconds or beyond \citep{DrorEtAl_PNAS11_DrugBindingGPCR,ShuklaPande_NatCommun14_SrcKinase,PlattnerNoe_NatComm15_TrypsinPlasticity,PlattnerEtAl_NatChem17_BarBar,PaulEtAl_PNAS17_Mdm2PMI}.
However, the extensive sampling of large macromolecular complexes
on biological timescales at atomistic resolution is still out of reach.
For this reason, the design of simplified, yet predictive models is
of great interest \citep{ClementiCOSB,Saunders2013,Noid2013}, in
particular, to interpret the experimental data that are becoming increasingly
accessible in high throughput and resolution. Experimental data provide
a partial view of certain aspects of a macromolecular system but do
not directly give a full dynamical representation and simulation can
help obtain a more comprehensive understanding \citep{MatysiakClementi_JMB04_Perturbation,MatysiakClementi_JMB06_Perturbation,Chen2018}.
As it is clear that not every single atom is important in determining
the relevant collective features of biomolecular dynamics and function,
simplified models could provide more insights into the general physicochemical
principles regulating biophysical systems at the molecular level.
Here we use recent advances in machine learning to design optimal
reduced models to reproduce the equilibrium thermodynamics of a macromolecule.

Significant effort has been devoted in the last few years to apply
machine learning (e.g., deep neural network or kernel methods) to
learn effective models from detailed simulations \citep{MardtEtAl_VAMPnets,WuEtAl_NIPS18_DeepGenMSM,WehmeyerNoe_TAE,HernandezPande_VariationalDynamicsEncoder,RibeiroTiwary_JCP18_RAVE},
and specifically to learn potential energy surfaces from quantum mechanical
calculations on small molecules \citep{BehlerParrinello_PRL07_NeuralNetwork,Bartok2010,Rupp2012,Bartok2013,Smith2017,ChmielaEtAl_SciAdv17_EnergyConserving,Bartok2017,Schuett2017,Smith2018,Schuett2018,Grisafi2018,Imbalzano2018,Nguyen2018,ZhangHan2018,ZhangHan2018_PRL,BereauEtAl_JCP18_Physical,Yang2019}.
In principle a similar philosophy could be used to define models at
lower resolutions, that is to learn the effective potential energy
of coarse-grained (CG) models from fine-grained (e.g., atomistic)
molecular dynamics (MD) simulation data \citep{John2017,ZhangHan2018_CG,ChanEtAl_NatComm19_CGWater,Peter2017,Deshmukh2018}.

There are however important differences. In the definition of potential
energy surfaces from quantum calculations, the relevant quantity to
reproduce is the energy, and it is relatively straightforward to design
a loss function for a neural network to minimize the difference between
the quantum mechanical and classical energy (and forces \citep{ChmielaEtAl_SciAdv17_EnergyConserving,ZhangHan2018})
over a sample of configurations. In contrast, in the definition of
a CG model, the potential energy can not be matched because of the
reduction in dimension, and it is important to define what are the
properties of the system that need to be preserved by the coarse-graining.
The approximation of free-energy surfaces, e.g. from enhanced sampling
simulations, is therefore a related problem \citep{StecherBernsteinCsanyi_JCTC14_FreeEnergyReconstruction,SchneiderEtAl_PRL17_FreeEnergyLearning,Jonathan2018}.

Several approaches have been proposed to design effective CG energy
functions for large molecular systems that either reproduce structural
features of atomistic models (bottom-up) \citep{Lyubartsev1995,MllerPlathe2002,Praprotnik2008,Izvekov2005,Wang2009,Shell2008}
or reproduce macroscopic properties for one or a range of systems
(top-down) \citep{ClementiJMB2000,Nielsen2003,MatysiakClementi_JMB04_Perturbation,MatysiakClementi_JMB06_Perturbation,Marrink2004,Davtyan2012,Chen2018}.
Popular bottom-up approaches choose that the CG model reproduce the
canonical configuration distribution determined by the atomistic model.
For instance, one may want to be able to represent the different metastable
states populated by a protein undergoing large conformational changes.
One of the difficulties in the practical application of these methods
has been that, in general, a CG potential optimally reproducing selected
properties of a macromolecular system includes many-body terms that
are not easily modeled in the energy functions.

Here, we formulate the well-known force matching procedure
for coarse-graining as a supervised machine learning problem. Previously,
coarse-graining has been mostly discussed as a fitting procedure,
but the aim of machine learning is to find a model that has minimal
prediction error on data not used for the training. We use classical
statistical learning theory to show that the force matching error
can be decomposed into Bias, Variance and Noise terms and explain
their physical meaning. We also show that the different CG models
can be ranked using their cross-validation score.

Second, we discuss a class of neural networks, which we refer to as CGnets, for
coarse-graining molecular force systems.  CGnets have a lot of similarities with
neural networks used to learn potential energy surfaces from quantum
data, such as enforcing the relevant invariances 
(e.g., rotational and translational invariance of the predicted energy, equivariance
of the predicted force). In contrast to potential energy networks,
CGnets predict a free energy (potential of mean force) and then use
the gradient of this free energy with respect to the input coordinates
to compute a mean force on the CG coordinates. As the CG free energy
is not known initially, only the force information can be used to
train the network.

Third, CGnets are extended to regularized CGnets. Using a generic
function approximator such as a neural network to fit the CG force
field from training data only may lead to force predictions that are
``catastrophically wrong'' for configurations not captured by the
training data, i.e., predictions of forces in the direction of increasingly
unphysical states that lead to diverging and unrealistic simulation
results. We address this problem by adding a prior energy to the free
energy network that does not compromise the model accuracy within
the training data region, but ensures that the free energy approaches
infinity for unphysical states, resulting in a restoring force towards
physically meaningful states.

Finally, we demonstrate that CGnets succeed in learning the CG mean
force and the CG free energy for a 2D toy model, as well as for the
coarse-graining of all-atom explicit-solvent simulations of (i) alanine
dipeptide to a CG model with 5 particles and no solvent, and (ii)
the folding/unfolding of the polypeptide Chignolin to a CG model consisting
only of the protein $C_{\alpha}$ atoms and no solvent. We show explicitly
that CGnets achieve a systematically better performance than classical
CG approaches which construct the CG free energy as a sum of few-body
terms. In the case of the Chignolin protein, the classical few-body
model can not reproduce the folding/unfolding dynamics. On the contrary,
the inherently multi-body CGnet energy function approximates the all-atom
folding/unfolding landscape well and captures all free energy minima.
This study highlights the importance of machine learning and generic
function approximators in the CG problem.

\section{Theory and methods}

Here we introduce the main theoretical concepts and define the machine
learning problems involved in coarse-graining using the force matching
principle, and introduce CGnets and regularized CGnets. The more practically
inclined reader may skip to the Section ``CGnets: Learning CG force fields with neural networks''.

\subsection{Coarse-graining with thermodynamic consistency}

We first define what we mean by coarse-graining and which physical
properties shall be preserved in the coarse-grained model.

The starting point in the design of a molecular model with resolution
coarser than atomistic is the definition of the variables. The choice
of the coarse coordinates is usually made by replacing a group of
atoms by one effective particle. Because of the modularity of a protein
backbone or a DNA molecule, popular models coarse-grain a macromolecule
to a few interaction sites per residue or nucleotide, e.g., the $C_{\alpha}$
and $C_{\beta}$ atoms for a protein \citep{ClementiJMB2000,voth2008coarse,Monticelli2008,Davtyan2012}.
Alternative schemes have also been proposed for the partitioning of
the atoms into coarse-grained coordinates \citep{sinitskiy2012optimal,BoninsegnaBanish2018}.
In general, given a high-dimensional atomistic representation of the
system $\mathbf{r}\in\mathbb{R}^{3N}$, a CG representation is given
by a coordinate transformation to a lower-dimensional space:
\begin{equation}
\mathbf{x}=\xi(\mathbf{r})\in\mathbb{R}^{3n}\label{eq:CGmapping}
\end{equation}
with $n<N$. Here we assume that $\xi$ is linear, i.e. there is some
coarse-graining matrix $\Xi\in\mathbb{R}^{3n\times3N}$ that clusters
atoms to coarse-grained beads: $\mathbf{x}=\Xi\mathbf{r}$.

The aim is to learn a coarse-grained energy function $U(\mathbf{x};\boldsymbol{\theta})$
that will be used in conjunction with a dynamical model, e.g., Langevin
dynamics, to simulate the CG molecule. $\boldsymbol{\theta}$ are
the parameters of the coarse-grained model -- in classical CG approaches
these are parameters of the potential energy function, such as force
constants and partial charges, while here they denote the weights
of the neural network.

A common objective in coarse-graining methods is to preserve the equilibrium
distribution, i.e. the equilibrium distribution of the coarse-grained
model shall be as close as possible to the equilibrium distribution
of the atomistic model when mapped to the CG coordinates. We will
be using a simulation algorithm for the dynamics such that the system's
equilibrium distribution is identical to the Boltzmann distribution
of the employed potential $U$; therefore this objective can be achieved
by enforcing the thermodynamic consistency:
\begin{equation}
U(\mathbf{x};\boldsymbol{\theta})\equiv-k_{B}T\ln p^{CG}(\mathbf{x})+\mathrm{const},\label{eq:free-ene}
\end{equation}
where $k_{B}T$ is the thermal energy with Boltzmann constant $k_{B}$
and temperature $T$, and the probability distribution $p^{CG}(\mathbf{x})$
is the equilibrium distribution of the atomistic model, mapped to
the CG coordinates:
\begin{equation}
p^{CG}(\mathbf{x}) = \frac{\int\mu(\mathbf{r})\delta\left(\mathbf{x} - 
\xi(\mathbf{r})\right)d\mathbf{r}}{\int\mu(\mathbf{r})d\mathbf{r}}
\label{eq:prob}
\end{equation}
and $\mu(\mathbf{r})=\exp\left(-V(\mathbf{r})/k_{B}T\right)$ is the
Boltzmann weight associated with the atomistic energy model $V(\mathbf{r})$.
Note that the additive constant in (\ref{eq:free-ene}) can be chosen
arbitrarily. Therefore this constant will be omitted in the expressions
below, which means that it will absorb normalization constants that
are not affecting the CG procedure, such as the logarithm of the partition
function.

Several methods have been proposed for defining a coarse-grained potential
$U(\mathbf{x})$ that variationally approximates the consistency relation
(\ref{eq:prob}) at a particular thermodynamic state (temperature,
pressure etc.) Two popular approaches are the multi-scale coarse-graining
(force-matching) \citep{Izvekov2005,Noid2008} and the relative entropy
method \citep{Shell2008} (the two approaches are connected \citep{Rudzinski2011}).

\subsection{CG parameter estimation as a machine learning problem}

Here, we follow the force-matching scheme. It has been shown that thermodynamic
consistency (\ref{eq:free-ene}) is achieved when the CG model predicts
the instantaneous CG forces with minimal mean square error \citep{Izvekov2005,Noid2008}.
We call the instantaneous atomistic forces $\mathbf{F}(\mathbf{r})$,
and the instantaneous force projected on the CG coordinates $\xi(\mathbf{F}(\mathbf{r}))$.
At the same time, the CG model predicts a force $-\nabla U(\mathbf{\mathbf{x}};\boldsymbol{\theta})$
for a CG configuration $\mathbf{\mathbf{x}}$. The force matching
error is defined as:
\begin{align}
\chi^{2}(\boldsymbol{\theta}) & =\left\langle \left\Vert \xi(\mathbf{F}(\mathbf{r}))+\nabla U(\xi(\mathbf{r});\boldsymbol{\theta})\right\Vert ^{2}\right\rangle _{\mathbf{r}}.\label{eq:force-matching}
\end{align}
The average $\langle\cdot\rangle_{\mathbf{r}}$ is over the equilibrium
distribution of the atomistic model, i.e., $\mathbf{r}\sim\mu(\mathbf{r})$.

We reiterate a result shown in \citep{Noid2008} that has important
consequences for using (\ref{eq:force-matching}) in machine learning.
For this, we introduce the mean force:
\begin{equation}
\mathbf{f}(\mathbf{x})=\langle\xi(\mathbf{F}(\mathbf{r}))\rangle_{\mathbf{r}\mid\mathbf{x}}\label{eq:mean_force}
\end{equation}
where $\mathbf{r}\mid\mathbf{x}$ indicates the equilibrium distribution
of $\mathbf{r}$ constrained to the CG coordinates $\mathbf{x}$,
i.e. the ensemble of all atomistic configurations that map to the
same CG configuration. Then we can decompose expression (\ref{eq:force-matching})
as follows (see SI for derivation):
\begin{align}
\chi^{2}(\boldsymbol{\theta}) & =\mathrm{PMF}\:\mathrm{error}(\boldsymbol{\theta})+\mathrm{Noise}\label{eq:force-matching-decomp1}
\end{align}
with the terms
\begin{align}
\mathrm{PMF}\:\mathrm{error}(\boldsymbol{\theta}) & =\langle\|\mathbf{f}(\xi(\mathbf{r}))+\nabla U(\xi(\mathbf{r});\boldsymbol{\theta})\|^{2}\rangle_{\mathbf{r}}\nonumber \\
\mathrm{Noise} & =\langle\|\xi(\mathbf{F}(\mathbf{r}))-\mathbf{f}(\xi(\mathbf{r}))\|^{2}\rangle_{\mathbf{r}}.\label{eq:noise}
\end{align}
This loss function differs from the force matching loss function used
in the learning of force fields from quantum data by the $\mathrm{Noise}$
term. The $\mathrm{Noise}$ term is purely a function of the CG map
$\xi$ (and when training with finite simulation data also of the
dataset), and it cannot be changed by varying the parameters $\boldsymbol{\theta}$.
As a result, the total force matching error cannot be made zero but
it is bounded from below by $\chi^{2}\left(\boldsymbol{\theta}\right)\ge\mathrm{Noise}$
\citep{Noid2008}. On the contrary, when matching force fields from
quantum data, the error $\chi^{2}$ approaches zero for a sufficiently
powerful model. Physically, the $\mathrm{Noise}$ term arises from
the fact that instantaneous forces on the CG coordinates vary in the
different atomistic configurations associated with the same CG configuration.
Here, we call this term $\mathrm{Noise}$ as it corresponds to the
noise term known in statistical estimator theory for regression problems
\citep{Vapnik_IEEE99_StatisticalLearningTheory}.

The learning problem is now to find a CG model and its parameters
$\boldsymbol{\theta}$ that minimizes the $\mathrm{PMF}\:\mathrm{error}$
term. In order to obtain a physical interpretation, we apply (\ref{eq:CGmapping})
and write the average purely in CG coordinates:
\begin{align*}
\mathrm{PMF}\:\mathrm{error}(\boldsymbol{\theta}) & =\langle\|\mathbf{f}(\mathbf{x})+\nabla U(\mathbf{x};\boldsymbol{\theta})\|^{2}\rangle_{\mathbf{x}}\\
 & =\langle\|\mathbf{f}(\mathbf{x})-\hat{\mathbf{f}}(\mathbf{x};\boldsymbol{\theta})\|^{2}\rangle_{\mathbf{x}}
\end{align*}
This error term is the matching error between the mean force at the
CG coordinates, $\mathbf{f}(\mathbf{x})$ and the CG forces predicted
by the CG potential,
\begin{equation}
\hat{\mathbf{f}}(\mathbf{x};\boldsymbol{\theta})=-\nabla U(\mathbf{x};\boldsymbol{\theta}).\label{eq:PMF}
\end{equation}
Hence, the machine learning task is to find the free energy $U$ whose
negative derivatives best approximate the mean forces in Eq (\ref{eq:mean_force}),
and $U$ is thus called a potential of mean force (PMF). Eq. (\ref{eq:PMF})
implies that the mean force field $\hat{\mathbf{f}}$ is conservative,
as it is generated by the free energy $U(\mathbf{x})$.

Machine learning the CG model is complicated by two aspects: (i) As
the PMF error cannot be computed directly, its minimization in practice
is accomplished by minimizing the variational bound (\ref{eq:force-matching-decomp1}).
Thus, to learn $\mathbf{f}(\mathbf{x})$ accurately, we need to collect
enough data ``close'' to every CG configuration $\mathbf{x}$ such
that the learning problem is dominated by the variations in the $\mathrm{PMF}\:\mathrm{error}$
term and not by the variations in the $\mathrm{Noise}$ term. As a
result, machine learning CG models typically requires more data points
than force matching for potential energy surfaces; (ii) The free energy
$U(\mathbf{x})$ is not known a priori, but must be learned. In contrast
to fitting potential energy surfaces we can therefore not directly
use energies as inputs.

For a finite dataset $\mathbf{R}=(\mathbf{r}_{1},...,\mathbf{r}_{M})$
with $M$ samples, we define the force matching loss function by the
direct estimator:
\begin{align}
L(\boldsymbol{\theta};\mathbf{R}) & =\frac{1}{3Mn}\sum_{i=1}^{M}\left\Vert \xi(\mathbf{F}(\mathbf{r}_{i}))+\nabla U(\xi(\mathbf{r}_{i});\boldsymbol{\theta})\right\Vert ^{2}\label{eq:training_loss}\\
 & =\frac{1}{3Mn}\left\Vert \xi(\mathbf{F}(\mathbf{R}))+\nabla U(\xi(\mathbf{R});\boldsymbol{\theta})\right\Vert _{F}^{2}.
\end{align}
Where $\xi(\mathbf{R})=\left[\xi(\mathbf{r}_{1}),...,\xi(\mathbf{r}_{M})\right]^{\top}\in\mathbb{R}^{M\times3n}$
and $\xi(\mathbf{F}(\mathbf{R}))=\left[\xi(\mathbf{F}(\mathbf{r}_{1})),...,\xi(\mathbf{F}(\mathbf{r}_{M}))\right]^{\top}\in\mathbb{R}^{M\times3n}$
are data matrices of coarse-grained coordinates and coarse-grained
instantaneous forces that serve as an input to the learning method,
and $F$ denotes the Frobenius norm.

\subsection{CG hyper-parameter estimation as a machine learning problem}

While Eq. (\ref{eq:training_loss}) defines the training method, machine
learning is not simply about fitting parameters for a given dataset,
but rather about minimizing the expected prediction error (also called
``risk'') for data not used for training. This concept is important
in order to be able to select an optimal model, i.e. in order to choose
the hyper-parameters of the model, such as the type and number of
neurons and layers in a neural network, or even to distinguish between
different learning models such as a neural network and a spline model. 

Statistical estimator theory is the field that studies optimal prediction
errors \citep{Vapnik_IEEE99_StatisticalLearningTheory}. To compute
the prediction error, we perform the following thought experiment:
We consider a fixed set of CG configurations $\mathbf{X}=[\mathbf{x}_{1},...,\mathbf{x}_{M}]^{\top}$
at which we want to fit the mean forces. We assume that these configurations
have been generated by MD or MCMC such that the full atomistic configurations,
$\mathbf{R}=(\mathbf{r}_{1},...,\mathbf{r}_{M})$, are Boltzmann distributions
conditioned on the CG configurations, i.e. $\mathbf{r}_{i}\sim\mathbf{r}\mid\mathbf{x}_{i}$.
Now we ask: if we repeat this experiment, i.e. in every iteration
we produce a new set of all-atom configurations $\mathbf{r}_{i}\sim\mathbf{r}\mid\mathbf{x}_{i}$,
and thereby a new set of instantaneous forces on the CG configurations,
what is the expected prediction error, or risk of the force matching
error, $\mathbb{E}\left[L(\boldsymbol{\theta};\mathbf{R})\right]$?
More formally:
\begin{enumerate}
\item Given CG coordinates $\mathbf{X}$, generate training set $\mathbf{R}^{\mathrm{train}}\sim\mathbf{R}\mid\mathbf{X}$
and find $\hat{\boldsymbol{\theta}}=\arg\min_{\boldsymbol{\theta}}L(\boldsymbol{\theta};\mathbf{R}^{\mathrm{train}})$.
\item Generate test set $\mathbf{R}^{\mathrm{test}}\sim\mathbf{R}\mid\mathbf{X}$
and compute $L(\hat{\boldsymbol{\theta}};\mathbf{R}^{\mathrm{test}})$
\end{enumerate}
where $\mathbf{R}^{\mathrm{train}}$ and $\mathbf{R}^{\mathrm{test}}$
are two independent realizations. Although we cannot execute this
thought experiment in practice, we can approximate it by cross-validation,
and we can obtain insightful expressions for the form of the expected
prediction error. As the loss function in force matching is a least
squares regression problem, the form of the expected prediction error
is well known (see SI for a short derivation), and can be written
as:
\begin{equation}
\mathbb{E}\left[L(\boldsymbol{\theta};\mathbf{R})\right]=\mathrm{Bias}^{2}+\mathrm{Var}+\mathrm{Noise}\label{eq:bias_variance_noise}
\end{equation}
with the $\mathrm{Noise}$ term as given in Eq. (\ref{eq:noise})
and the bias and variance terms given by:
\begin{align}
\mathrm{Bias}^{2} & =\left\Vert \mathbf{f}(\mathbf{\mathbf{X}})-\bar{\mathbf{f}}(\mathbf{\mathbf{X}})\right\Vert _{F}^{2}\label{eq:bias}\\
\mathrm{Var} & =\mathbb{E}\left[\left\Vert \bar{\mathbf{f}}(\mathbf{\mathbf{X}})+\nabla U(\mathbf{X})\right\Vert _{F}^{2}\right]\label{eq:variance}
\end{align}
where
\[
\bar{\mathbf{f}}(\mathbf{X})=\mathbb{E}\left[-\nabla U(\mathbf{X})\right]
\]
is the mean estimator, i.e. the average force field learnt when the
training is repeated many times for different data realizations. The
terms in (\ref{eq:bias}-\ref{eq:variance}) have the following meaning:
Eq. (\ref{eq:bias}) is the expected error between the mean forces
and the average predicted force field, it is therefore the systematic
bias of the machine learning model. The variance (\ref{eq:variance})
is the fluctuation of the individual estimates from single training
procedures around the mean estimator and thus represents the estimator's
fluctuation due to finite-sample effects.

\begin{figure}[h]
\begin{centering}
\includegraphics[width=0.8\columnwidth]{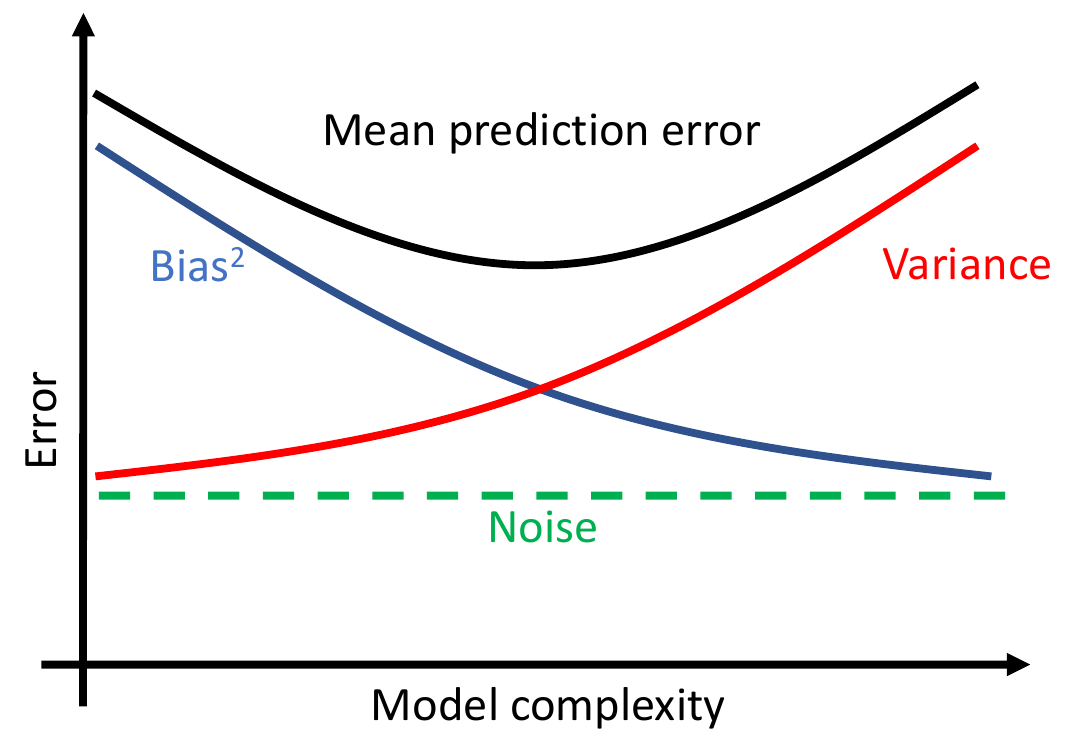}
\par\end{centering}
\caption{\footnotesize\label{fig:bias_variance}Typical bias-variance tradeoff for fixed
data set size, indicating the balance between underfitting and overfitting.
The noise level is defined by the CG scheme (i.e., which particles
are kept and which are discarded) and is the lower bound for the mean
prediction error.}
\end{figure}

As the optimal model minimizes the PMF error, it must balance bias
and variance. These contributions are typically counteracting: A too
simple model (e.g., too small neural network) typically leads to low
variance but high bias, and it corresponds to ``underfitting'' the
data. A too complex model (e.g., too large neural network) leads to
low bias but large variance, and it corresponds to ``overfitting''
the data. The behavior of bias, variance and estimator error for a
fixed data set size is illustrated in Fig. \ref{fig:bias_variance}.

The optimum at which bias and variance balance depends on the amount
of data used, and in the limit of an infinitely large dataset, the
variance is zero and the optimal model can be made very complex so
as to also make the bias zero. For small datasets, it is often favorable
to reduce the model complexity and accept significant bias, in order
to avoid large variance.

In order to implement model selection, we approximate the ``ideal''
iteration above by the commonly used cross-validation method \citep{DevijverKittler_PatternRecognitionBook,Kohavi_CrossValidation}
and then choose the model or hyper-parameter set that has the minimal
cross-validation score. In cross-validation, the estimator error (\ref{eq:bias_variance_noise})
is estimated as the validation error, averaged over different segmentations
of all available data into training and validation data.

\subsection{CGnets: Learning CG force fields with neural networks}

\label{subsec:CGnets}

It is well known that the CG potential $U(\mathbf{x};\boldsymbol{\theta})$
defined by thermodynamic consistency may be a complex multi-body potential
even if the underlying atomistic potential has only few-body interactions
\citep{Noid2008}. To address this problem, we use artificial neural
networks (ANNs) to represent $U(\mathbf{x};\boldsymbol{\theta})$
as ANNs can approximate any smooth function on a bounded set of inputs,
including multi-body functions \citep{HornikEtAlNeuralNetworks91UniversalApproximation}.

\begin{figure}[h]
\begin{centering}
\includegraphics[width=1\columnwidth]{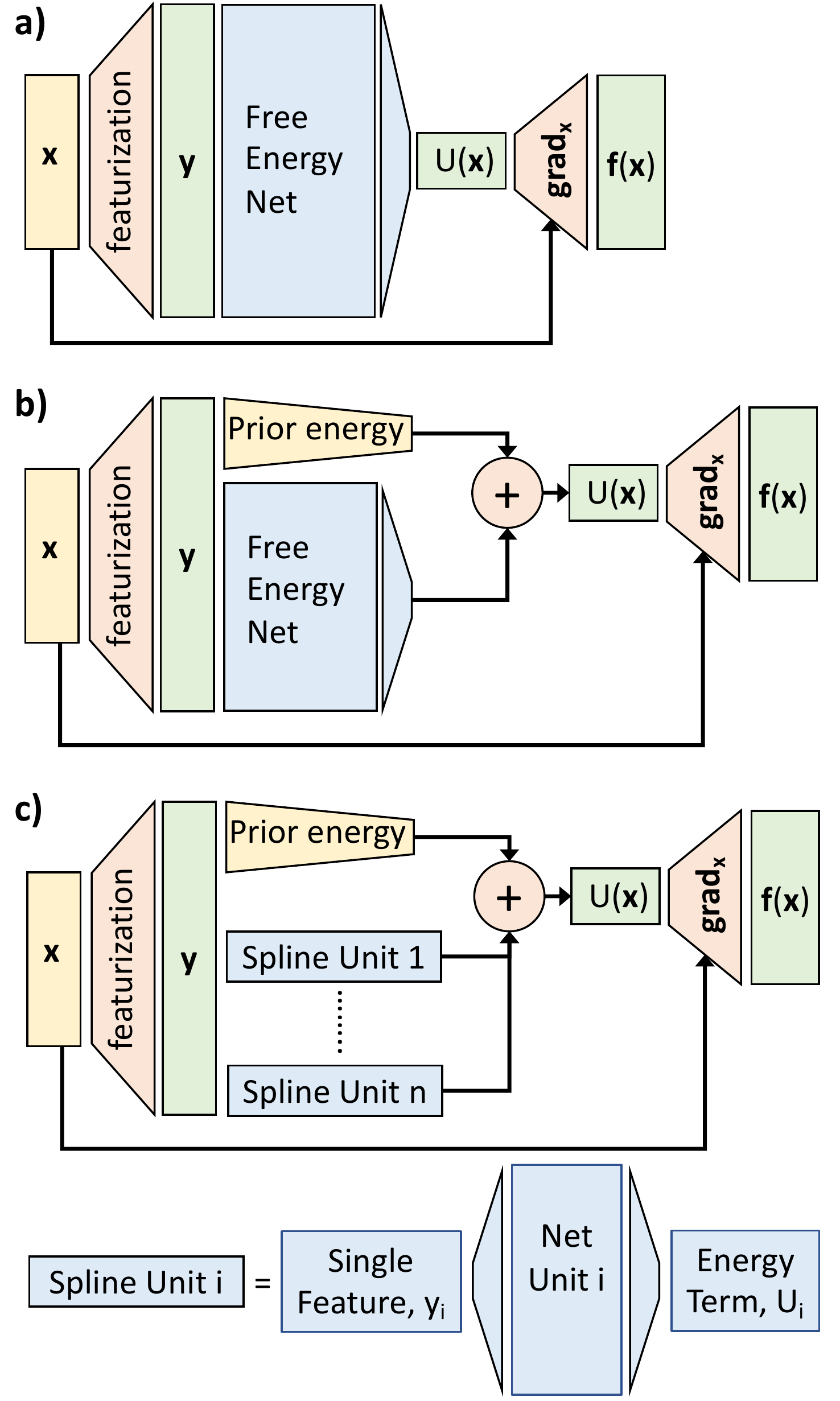}
\par\end{centering}
\caption{\label{fig:network_scheme}
\footnotesize Neural network schemes. a) CGnet. b) Regularized
CGnet with prior energy. c) Spline model representing a standard CG
approach, for comparison. Each energy term is a function of only one
feature, and the features are defined as all the bonds, angles, dihedrals,
and non-bonded pairs of atoms.}
\end{figure}

Therefore, we use ANNs to model $U(\mathbf{x})$, train them by minimizing
the loss (\ref{eq:training_loss}) and select optimal models by minimizing
the cross-validation loss. For the purpose of training CG molecular
models, we would like to have the following physical constraints and
invariances, which determine parts of the architecture of the neural
network:
\begin{itemize}
\item \textbf{Differentiable free energy function}: In order to train $U(\mathbf{x};\boldsymbol{\theta})$
and simulate the associated dynamics by means of Langevin simulations,
it must be continuously differentiable. As the present networks do
not need to be very deep, vanishing gradients are not an issue and
we select $\tanh$ activation functions here. After $D$ nonlinear
layers we always add one linear layer to map to one output neuron
representing the free energy.
\item \textbf{Invariances of the free energy}: The energy of molecules that
are not subject to an external field only depends on internal interactions
and is invariant with respect to translation or rotation of the entire
molecule. The CG free energy may also be invariant with respect to
permutation of certain groups of CG particles, e.g. exchange of identical
molecules, or certain symmetric groups within molecules. Compared
to quantum-mechanical potential energies, permutation invariance is
less abundant in CG. For example, permutation invariance does not
apply to the $\alpha$-carbons in a protein backbone (not even for
identical amino acids), as they are ordered by the MD bonding topology.
CGnets include a transformation:
\[
\mathbf{y}=g(\mathbf{x})
\]
from CG Cartesian coordinates $\mathbf{x}$ to a set of features that
contain all desired invariances, and use the features $\mathbf{y}$
as an input to the network that computes the free energy, $U(g(\mathbf{x});\boldsymbol{\theta})$.
This transformation can be chosen in many different ways, e.g. by
using local coordinate systems \citep{ZhangHan2018_PRL}, two- or
three-body correlation functions \citep{BehlerParrinello_PRL07_NeuralNetwork},
permutation-invariant distance metrics \citep{HansenEtAl_JCTC13_AssessmentFeaturization,BartokKondorCsanyi_PRB13_SOAP,ChmielaEtAl_NatComm18_TowardExact},
or by a learned representation \citep{Schuett2018}. In this work, only
translation and rotation invariances are needed, and we
hence choose the following features: distances between all pairs of
CG atoms, the angles between three consecutive CG atoms, and the $cos$
and $sin$ of torsion angles defined by the CG atoms.
\item \textbf{Conservative PMF}: The PMF is a conservative force field generated
by the free energy (\ref{eq:PMF}). As in quantum potential energy
learning \citep{ChmielaEtAl_SciAdv17_EnergyConserving,Schuett2018}, we enforce this requirement
by computing the free energy $U$ with a neural network and then adding
a gradient layer to compute the derivatives with respect to the input
coordinates:
\[
\hat{\mathbf{f}}(\mathbf{x};\boldsymbol{\theta})=-\nabla_{\mathbf{x}}U(g(\mathbf{x});\boldsymbol{\theta}).
\]
\end{itemize}
Fig. \ref{fig:network_scheme}a shows the neural network architecture
resulting from these choices. The free energy network is $D$ layers
deep and each layer is $W$ neurons wide. Additionally, we use L2 Lipschitz
regularization \citep{Lipschitz} in the network, with a tunable parameter
$\lambda$ that regulates the strength of the regularization. Thus,
$(D,W,\lambda)$ are the remaining hyper-parameters to be selected
(as discussed in the Results section).

\subsection{Simulating the CGnet model}

Once the neural network has been trained to produce a free energy
$U(\mathbf{x})$, it can be used to simulate dynamical trajectories
of the CG model. Here we use over-damped Langevin dynamics to advance
the coordinates of the CG model from $\mathbf{x}_{t}$ at time $t$
to $\mathbf{x}_{t+\tau}$ after a time-step $\tau$ :

\begin{equation}
\mathbf{x}_{t+\tau}=\mathbf{x}_{t}-\tau\frac{D}{k_{B}T}\nabla U(\mathbf{x}_{t})+\sqrt{2\tau D}\boldsymbol{\xi}\label{eq:Smoluchowski}
\end{equation}
where $\mathbf{x}_{t}$ is the CG configuration at time $t$ (e.g.,
the $x$ coordinate in the toy model, a 15-dimensional vector in
the alanine dipeptide, and a 30-dimensional vector in the Chignolin
applications presented below). $\boldsymbol{\xi}$ is Gaussian random
noise with zero mean and identity as covariance matrix, $\tau$ is
the integration time step, $D$ is the diffusion constant of the system.
In the following, we use reduced energy units, i.e. all energies are
in multiples of $k_{B}T$.

Since the implementation of CGnet is vectorized, it is more efficient
to compute free energies and mean forces for an entire batch of configurations,
rather than a single configuration at a time. Therefore, we run simulations
in parallel for the examples shown below. This is done by sampling
100 starting points randomly from atomistic simulations, coarse-graining
them and then integrating (\ref{eq:Smoluchowski}) stepwise.

\subsection{Regularizing the free energy with a baseline energy model}
Training the free energy with a network as shown in Fig. \ref{fig:network_scheme}a
and subsequently using it in order to simulate the dynamics with Eq.
(\ref{eq:Smoluchowski}) produces trajectories of new CG coordinates
$\mathbf{x}_{t}$. When parts of the coordinate space are reached
that are very different from any point in the training set, it is
possible that the network makes unphysical predictions.

In particular, the atomistic force-field used to produce the training
data has terms that ensure the energy will go towards infinity when
departing from physical states, e.g. when stretching bonds or when
moving atoms too close to each other. These regions will not be sampled
in the underlying MD simulations, and therefore result in ``empty''
parts of configuration space that contain no training data. Simulating
a network trained only on physically valid training data via Eq. (\ref{eq:Smoluchowski})
may still produce points $\mathbf{x}_{t}$ that enter this ``forbidden
regime'' where bonds are overstretched or atoms start to overlap.
At this point the simulation can become unstable if there is no regularizing
effect ensuring that the predicted free energy $U(\mathbf{x};\boldsymbol{\theta})$
will increase towards infinity when going deeper into the forbidden
regime.

Methods to modify a learning problem so as to reduce prediction errors
are collectively known as ``regularization'' methods \cite{GoodfellowBengioCourville_BookDL}.
In order to avoid the catastrophically wrong prediction problem described
above, we introduce regularized CGnets (Fig. \ref{fig:network_scheme}b).
In a regularized CGnet, we define the energy function as
\begin{equation}
U(\mathbf{x};\boldsymbol{\theta})=U_{0}(\mathbf{x})+U_{\mathrm{net}}(\mathbf{x};\boldsymbol{\theta})\label{eq:posterior_energy}
\end{equation}
where $U_{\mathrm{net}}(\mathbf{x};\boldsymbol{\theta})$ is a neural
network free energy as before and $U_{0}(\mathbf{x})$ is a baseline
energy that contains constraint terms that ensure basic physical behavior.
Such baseline energies to regularize a more complex multi-body energy
function have also been used in the machine learning of QM potential
energy functions \citep{Shapeev_MMS16_MomentTensorPotentials,DolgirevEtAl_APIAdv16_ChemicallyInterpretable,DeringerCsanyi_PRB17_MLCarbon}.
Note that (\ref{eq:posterior_energy}) can still be used to represent
any smooth free energy because $U_{\mathrm{net}}(\mathbf{x};\boldsymbol{\theta})$
is a universal approximator. The role of $U_{0}(\mathbf{x})$ is to
enforce $U\rightarrow\infty$ for unphysical states $\mathbf{x}$
that are outside the training data. 

As for many other regularizers, the baseline energy $U_{0}(\mathbf{x})$
in Eq. (\ref{eq:posterior_energy}) takes the role of a prior distribution
in a probabilistic interpretation: The equilibrium distribution generated
by (\ref{eq:posterior_energy}) becomes:
\[
p^{CG}(\mathbf{x})\propto\underset{\mathrm{prior}}{\underbrace{\exp\left(-\beta U_{0}(\mathbf{x})\right)}}\exp\left(-\beta U_{net}(\mathbf{x};\theta)\right).
\]

Here, $U_{0}(\mathbf{x})$ is
simply a sum of harmonic and excluded volume terms. For the 2d toy model, a harmonic
term in the form $U_{0}(x)=\frac{1}{2}k(x-x_{0})^{2}$ is used, and
the parameters $k$ and $x_{0}$ are determined by the force matching
scheme restricted to the scarcely populated regions defined by the
100 sampled points with highest and the 100 with lowest $x$-value
(see Fig. \ref{fig:2D-potential}).

For alanine dipeptide, we use harmonic terms for the distance between
atoms that are adjacent (connected by covalent bonds) and for angles
between three consecutive atoms. For each bond $i$, we use $U_{0,i}^{bond}(r_{i};r_{i0},k_{b,i})=\frac{1}{2}k_{b,i}(r_{i}-r_{i0})^{2}$,
where $r_{i}$ is the instantaneous distance between the two consecutive
atoms defining the bond, $r_{i0}$ is the equilibrium bond length,
and $k_{b,i}$ is a constant. Analogously, for each angle $j,$ we
use $U_{0,j}^{angle}(\theta_{j};\theta_{j0},k_{a,j})=\frac{1}{2}k_{a,j}(\theta_{j}-\theta_{j0})^{2}$,
where $\theta_{j}$ is the instantaneous value of the angle, $\theta_{j0}$
is the equilibrium value for the angle, and $k_{a,j}$ is a constant.
When statistically independent, each such term would give rise to
a Gaussian equilibrium distribution:
\begin{align*}
p(r_{i}) & \wasypropto\exp\left(-\frac{k_{b,i}(r_{i}-r_{i0})^{2}}{2k_{B}T}\right)\\
p(\theta_{j}) & \wasypropto\exp\left(-\frac{k_{a,j}(\theta_{j}-\theta_{j0})^{2}}{2k_{B}T}\right)
\end{align*}
with mean $\mu=r_{i0}$ (or $\mu=\theta_{j0}$), and variance $\sigma^{2}=k_{B}T/k_{b,i}$
(or $\sigma^{2}=k_{B}T/k_{a,j}$ ). The prior energy is obtained by
assuming independence between these energy terms and estimating these
means and variances from the atomistic simulations. 

For the application of CGnet to the protein Chignolin, an additional
term is added to the baseline energy to enforce excluded volume and
penalize clashes between non-bonded CG particles. In particular, we
add a term $U_{rep}(r)$ for each pairwise distances between CG particles
that are more distant than two covalent bonds, in the form:

\begin{equation}
U_{rep}(r)=\left(\frac{\sigma}{r}\right)^{c}\label{eq:excluded_volume}
\end{equation}

where the exponent $c$ and effective excluded volume radius $\sigma$
are two additional hyper-parameters that are optimized by cross-validation.

We note that in general one could use classical CG approaches with
predefined energy functions to first define the prior CG energy $U_{0}$
, then use an ANN to correct it with multi-body terms.

\section{Results}

\subsection{2-dimensional toy model}
\begin{figure}[!h]
\vspace{-0.5cm}
\begin{centering}
\includegraphics[width=0.8\columnwidth]{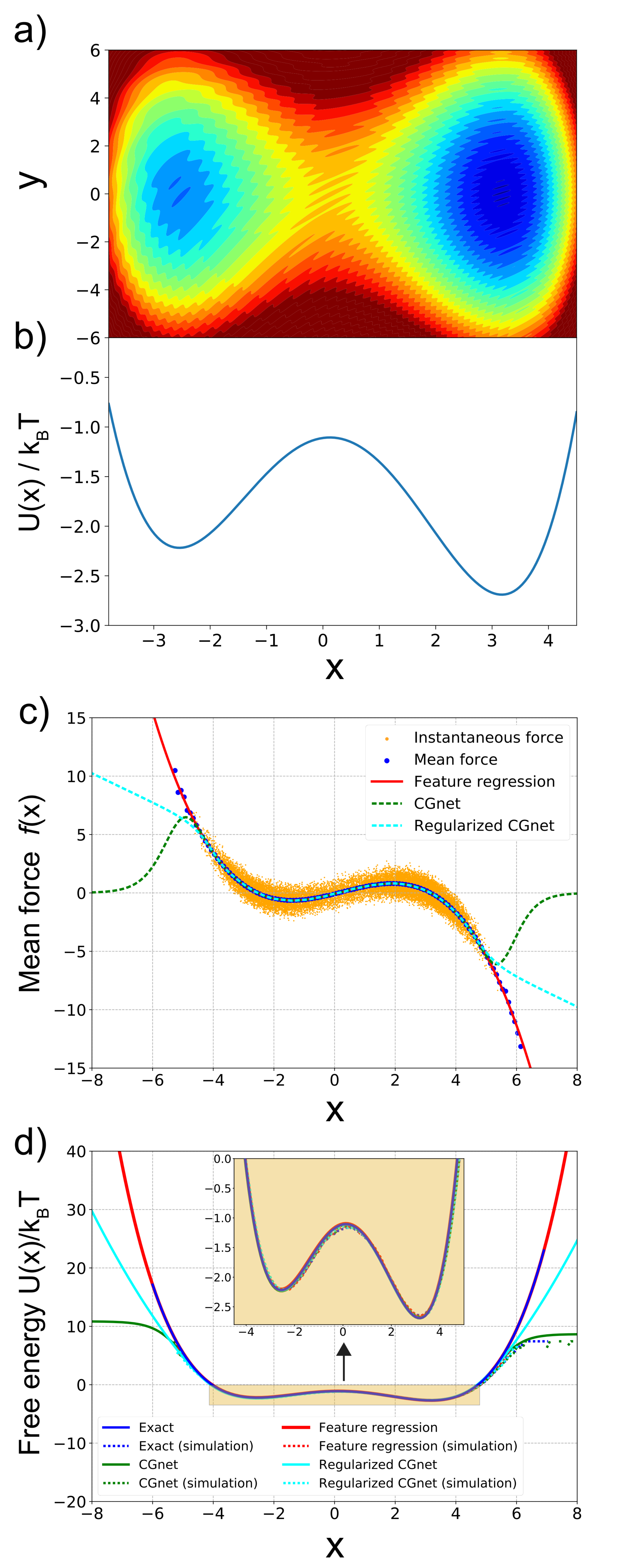}
\par\end{centering}
\caption{\label{fig:2D-potential}
\footnotesize Machine-learned coarse-graining of dynamics
in a rugged 2D potential. (a) 2D potential used as toy system. (b)
Exact free energy along $x$. c) Instantaneous forces and the learned
mean forces using feature regression and CGnet models (regularized
and unregularized) compared to the exact forces. The unit of the force
is $k_{B}T$, with the unit of length equal to 1. d) Free energy (PMF)
along $x$ predicted using least feature regression, and CGnet models
compared to the exact free energy. Free energies are also computed
from histogramming simulation data directly, using the underlying
2D trajectory, or simulations run with the feature regression and
CGnet models (dashed lines).}
\end{figure}

As a simple illustration, we first present the results on the coarse-graining
of a two-dimensional toy model. The potential energy is shown in Fig.
\ref{fig:2D-potential} and given by the expression:
\begin{multline}
\frac{V(x,y)}{k_{B}T}=\frac{1}{50}(x-4)(x-2)(x+2)(x+3)+\\
+\frac{1}{20}y^{2}+\frac{1}{25}\sin\left(3(x+5)(y-6)\right).\label{eq:2dpot}
\end{multline}
The potential corresponds to a double well along the $x$-axis and
a harmonic confinement along the $y$-axis. The last term in Eq. (\ref{eq:2dpot})
adds small-scale fluctuations, appearing as small ripples in Fig.
\ref{fig:2D-potential}a.

The coarse-graining mapping is given by the projection of the 2-dimensional
model onto the $x$-axis. In this simple toy-model, the coarse-grained
free energy (potential of mean force) can be computed exactly (Fig.
\ref{fig:2D-potential}b):
\begin{align*}
\frac{U(x)}{k_{B}T} & =-\ln\left[\int_{-\infty}^{+\infty}\exp\left(-\frac{V(x,y)}{k_{B}T}\right)dy\right].
\end{align*}

We generate a long (one million time steps) simulation
trajectory of the 2-dimensional model and use the $x$ component
of the forces computed along the trajectories in the loss function
(\ref{eq:training_loss}). We report below the resulting CG potential
obtained by 1) using a feature regression, i.e. least square regression
with a set of feature functions defined in SI Section B, and 2) a
CGnet (regularized and unregularized).

Cross-validation is used to select the best hyper-parameters for the
least square regression and the CGnet architectures. For the feature
regression, the same cross-validation procedure as introduced in \citep{Boninsegna2018}
was used, and returned a linear combination of four basis functions
among the selected set (see Fig. S1a and SI for details).
For the regularized CGnet, a two stage cross-validation is conducted,
first choosing the depth $D$ with a fixed width of $W=50$ , and
then choosing the width $W$ (Figs. S1b and S1c). The minimal
prediction error is obtained with $D=1$ (one hidden layer) and $W=50$.
For the unregularized CGnet, a similar procedure is performed, and
the best hyper-parameters are selected as $D=1,W=120$. Note that
the prediction error cannot become zero, but is bounded from below
by the chosen CG scheme (Fig. \ref{fig:bias_variance}, Eq. \ref{eq:bias_variance_noise})
-- in this case by neglecting the $y$ variable.

Fig. \ref{fig:2D-potential}c,d shows the results of the predicted
mean forces and free energies (potentials of mean force) in the $x$-direction.
The instantaneous force fluctuates around the mean, but serves to
accurately fit the exact mean force in the $x$ range where sampling
is abundant using both feature regression and CGnet (Fig. \ref{fig:2D-potential}c).
At the boundary where few samples are in the training data the predictors
start to diverge from the exact mean force and free energy (Fig. \ref{fig:2D-potential}c,
d). This effect is more dramatic for the unregularized CGnet, in particular
at large $x$ values the CGnet makes an arbitrary prediction: here
the force tends to zero. In the present example, reaching these states
is highly improbable. However a CGnet simulation reaching this region
can fail dramatically, as the simulation may continue to diffuse away
from the low energy regime. As discussed above, this behavior can
be avoided by adding a suitable prior energy that ensures that the
free energy keeps increasing outside the training data, while not
affecting the accuracy of the learned free energy within the training
data (Fig. \ref{fig:2D-potential}c, d). Note that the quantitative
mismatch in the low-probability regimes is not important for equilibrium
simulations.

The matching mean forces translate into matching free energies (potentials
of mean force, Fig. \ref{fig:2D-potential}d). Finally, we conduct
simulations with the learned models and generate trajectories $\{x_{t}\}$
using Eq. (\ref{eq:Smoluchowski}). From these, free energies can
be computed by
\begin{equation}
\tilde{U}(\mathbf{x})=-k_{B}T\ln\tilde{p}_{X}(\mathbf{x})\label{eq:free_energy_data}
\end{equation}
where $\tilde{p}_{X}(\mathbf{x})$ is a histogram estimate of the
probability density of $\mathbf{x}$ in the simulation trajectories.
As shown in Fig. \ref{fig:2D-potential}d, free energies agree well
in the $x$ range that has significant equilibrium probability.

\subsection{Coarse-graining of alanine dipeptide in water}

\begin{figure}[t]
\centering{}\includegraphics[width=0.8\columnwidth]{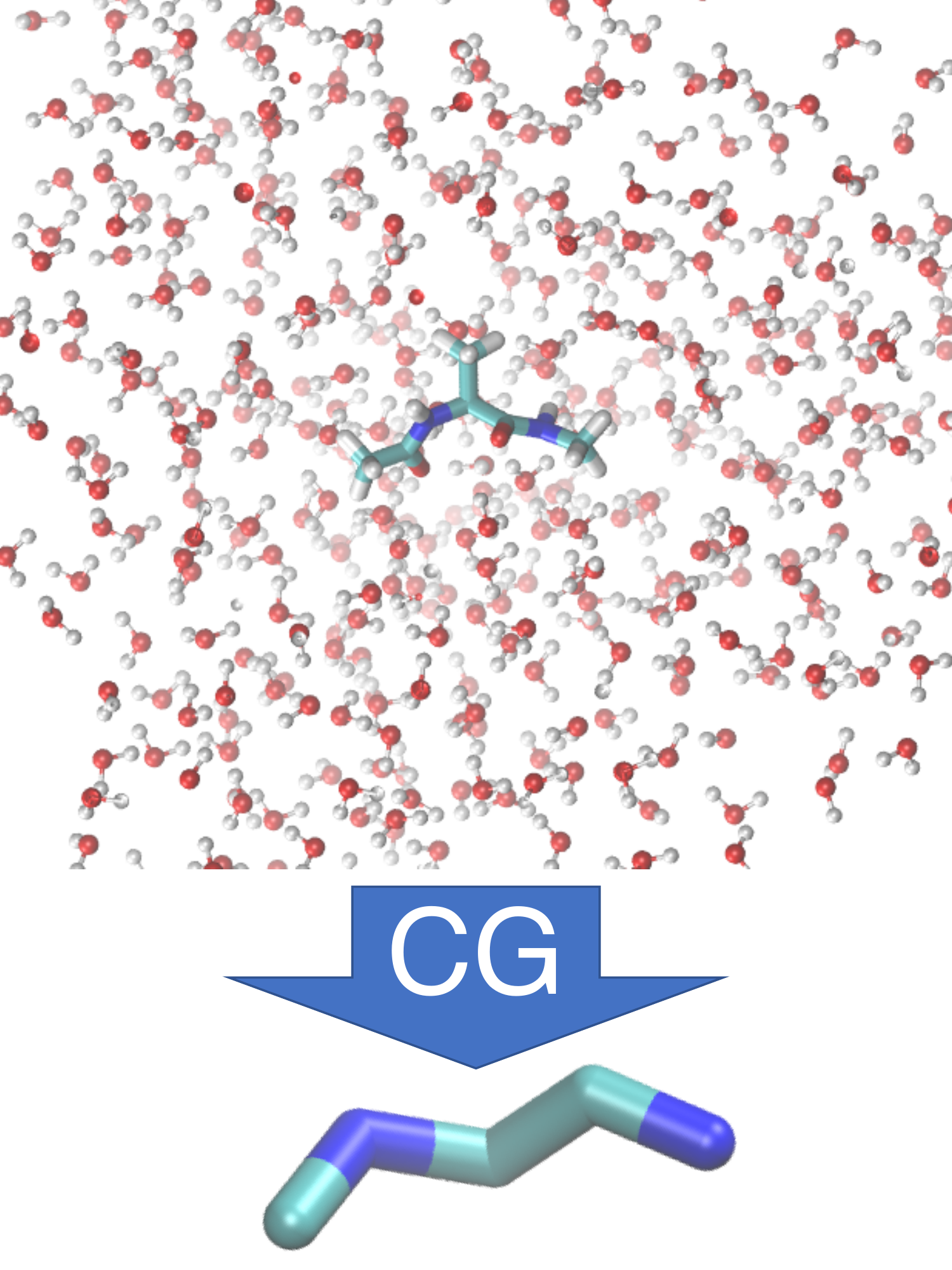}
\caption{\label{fig:AlaDipMapping}
\footnotesize
Mapping of alanine dipeptide from an all-atom
solvated model (top) to a CG model consisting of the five central
backbone atoms (bottom).}
\end{figure}

\begin{figure}[!th]
\centering{}\includegraphics[width=0.99\columnwidth]{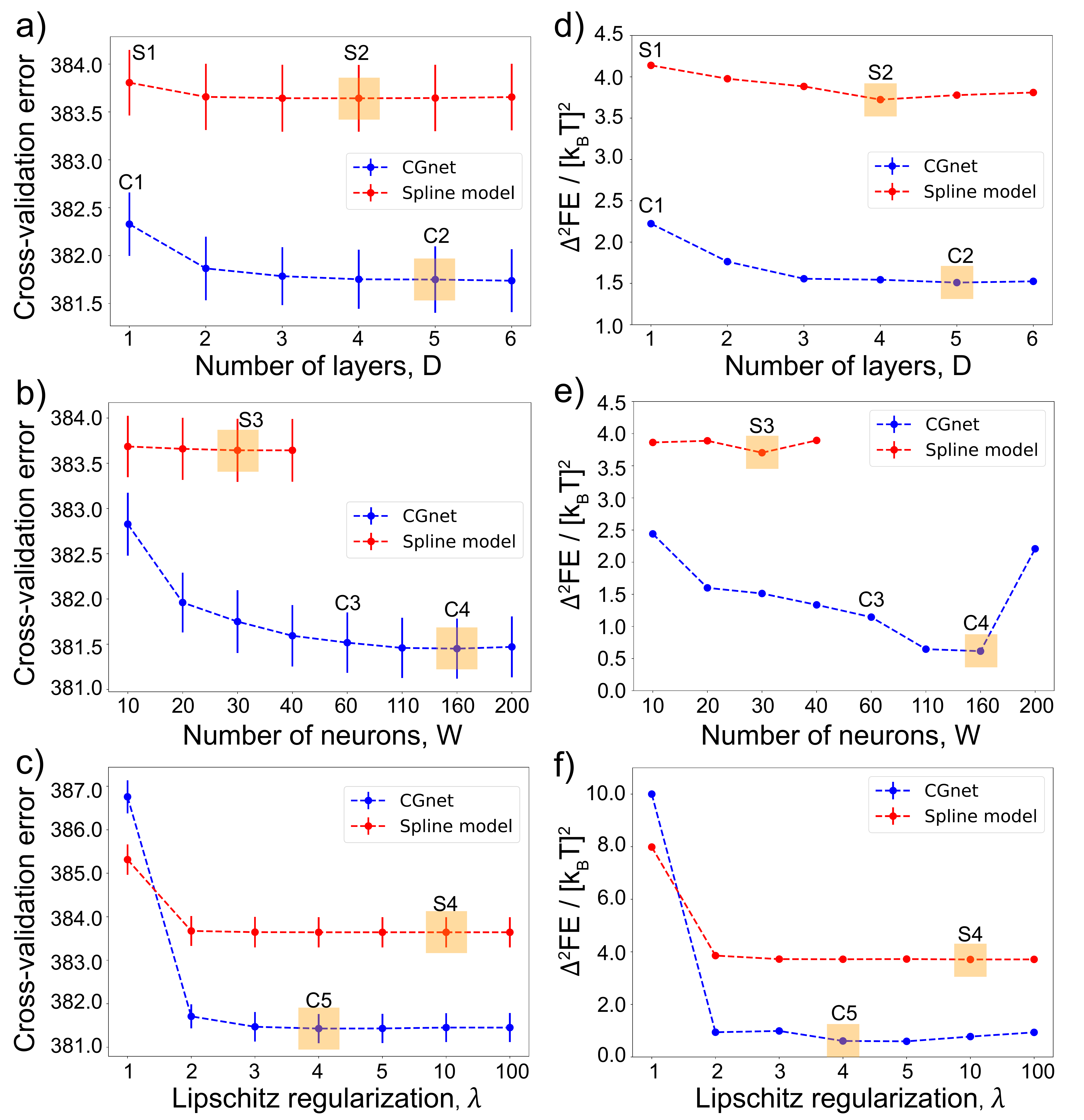}
\caption{\label{fig:crossvalidation-1}
\footnotesize 
(a), (b), (c) Cross-validated force matching error
in $[kcal/(mol\mathring{\cdot\textrm{A}})]^{2}$ for the selection
of the optimum structure of the network, and (d), (e), (f) difference between the two-dimensional
free energy surfaces obtained from the CG models and from the reference
all-atom simulations (see Fig. \ref{fig:Free-energy})
 for the regularized CGnet and the spline model of alanine dipeptide. 
(a) Selection of the number of layers, $D$.
(b) Selection of the number of neurons per layer, $W$. (c) Selection
of the Lipschitz regularization strength, $\lambda$. The selected
hyper-parameters, corresponding to the smallest cross-validation error
are highlighted by orange boxes. Blue dashed lines represent the regularized
CGnet, red dashed lines represent the spline model, vertical bars
represent the standard error of the mean. Panels (d), (e), and (f) show the
difference between the reference all-atom free energy surface and the free
energy surfaces corresponding to the choices of hyper-parameters indicated in
panels (a), (b), and (c) as (C1, C2, C3, C4, C5) for CGnet, and as (S1, S2, S3,
S4) for the spline model.} 
\end{figure}

We now demonstrate CGnets on the coarse-graining of an all-atom MD
simulation of alanine dipeptide in explicit solvent at $T=300K$ to
a simple model with 5 CG particles located at the five central backbone
atoms of the molecule (Fig. \ref{fig:AlaDipMapping}). One trajectory
of length $1$ microsecond was generated using the simulation setup
described in \citep{NueskeEtAl_JCP17_OOMMSM}, coordinates and forces
were saved every picosecond, giving rise to one million data points.
The CG model has no solvent, therefore the CG procedure must learn
the solvation free energy for all CG configurations.

\begin{figure}[!h]
\begin{centering}
\includegraphics[width=1\columnwidth]{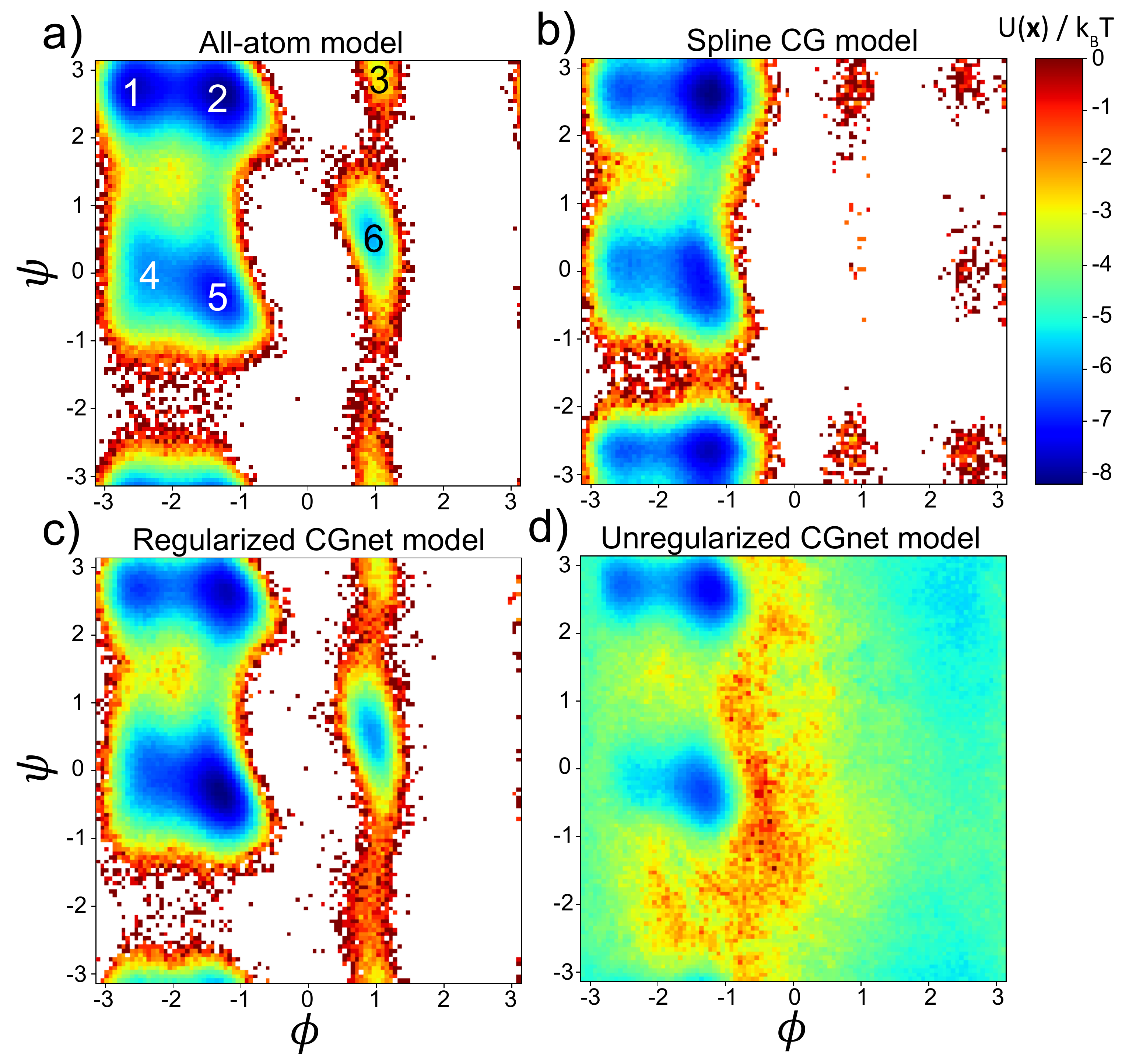}
\par\end{centering}
\begin{centering}
\includegraphics[width=1\columnwidth]{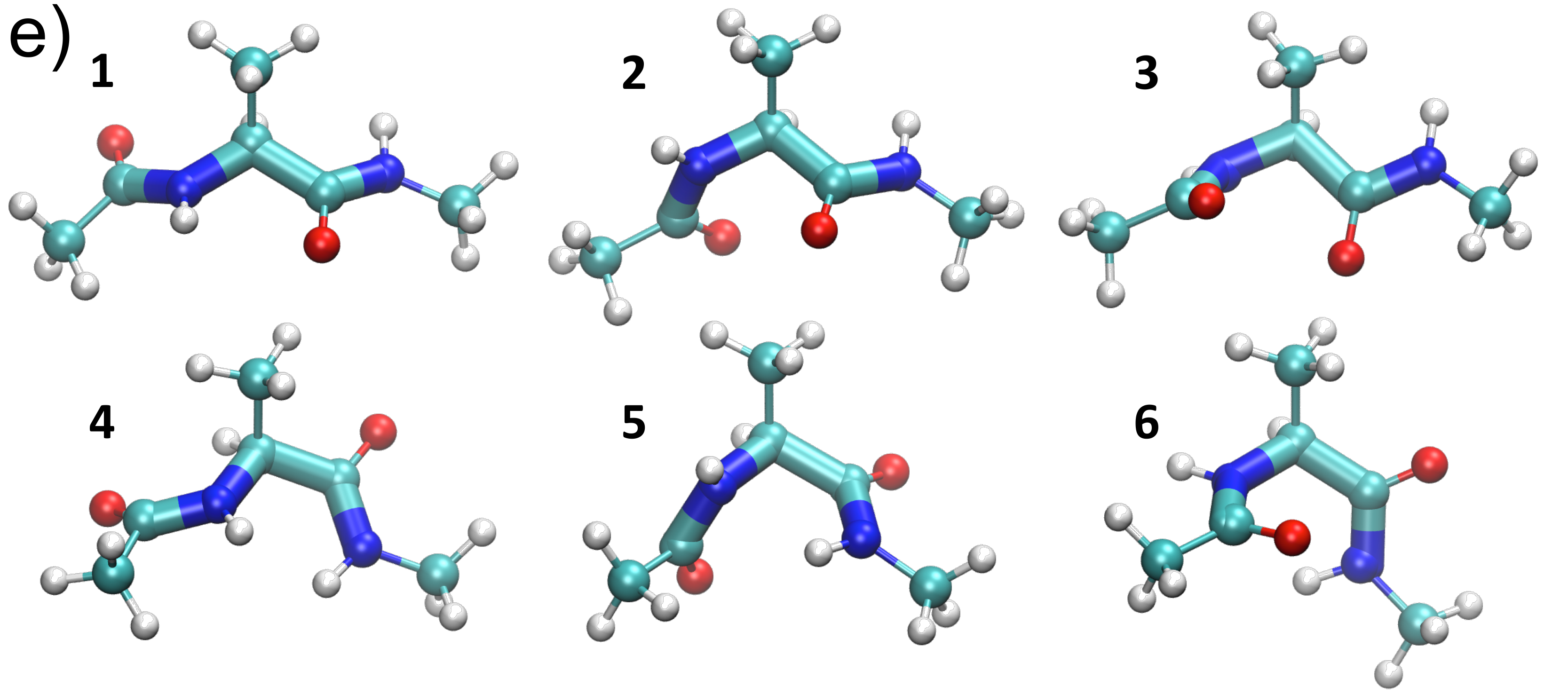}
\par\end{centering}
\caption{\label{fig:Free-energy}
\footnotesize
Free energy profiles and simulated structures
of alanine dipeptide using all-atom and machine-learned coarse-grained
models. (a)Reference free energy as a function of the dihedral angles,
as obtained from direct histogram estimation from all-atom simulation.
(b) Standard coarse-grained model using a sum of splines of individual
internal coordinates. (c) Regularized CGnet as proposed here. (d)
Unregularized CGnet. (e) Representative structures in the six free
energy minima, from atomistic simulation (ball-and-stick representation)
and regularized CGnet simulation (licorice representation).}
\end{figure}

We compare two different CG models. The first model, called ``spline
model'', uses the state-of the art approach established in MD coarse-graining
\citep{Noid2008,Wang2009,Noid2013}: to express the CG potential as
a sum of few-body interaction terms, similar as in classical MD forcefields.
The most flexible amongst these approaches is to fit one-dimensional
splines for each of the pairwise distance, angle and dihedral terms
in order to parametrize two-, three- and four-body interactions \citep{Dunn2017}.
In order to ensure a consistent comparison, we represent 1D splines
with neural networks that map from a single input feature (pairwise
distance, angle or dihedral) to a single free energy term, resulting
in the spline model network shown in Fig. \ref{fig:network_scheme}c.
We use the same regularization and baseline energy for spline model
networks and CGnets.

The second model uses a regularized multi-body CGnet, i.e., a fully
connected neural network shown in Fig. \ref{fig:network_scheme}b
to approximate the CG free energy. The comparison of the results from
the two models allows us to evaluate the importance of multi-body interactions
that are captured by the CGnet but are generally absent in CG models
that use interaction terms involving a few atoms only.

The hyper-parameters for both models consist of the number of layers
(depth, $D$), the number of neurons per layer (width, $W$) of the
network, and the Lipschitz regularization strength ($\lambda$) \citep{Lipschitz},
and are optimized by a three-stage cross-validation. In the first
stage, we find the optimal $D$ at fixed $W=30$ and $\lambda=\infty$
(no Lipschitz regularization), subsequently we choose $W$ at the
optimal $D$, and $\lambda$ at the optimal $W,D$. This results in
$D=5$ , $W=160$, and $\lambda=4.0$ for CGnet, and $D=4$ , $W=30$ (for each feature),
and $\lambda=10.0$  for the spline model (Fig. \ref{fig:crossvalidation-1}). The cross-validation error of CGnet is significantly
lower than the cross-validation error of the spline model (Fig. \ref{fig:crossvalidation-1}a-c).
We point out that the cross-validation error cannot become zero, but
is bounded from below by the chosen CG scheme (Fig. \ref{fig:bias_variance},
Eq. \ref{eq:bias_variance_noise}) -- in this case by coarse-graining
all solvent molecules and all solute atoms except the five central
backbone atoms away. Hence, the absolute numbers of the cross-validation
error in Fig. \ref{fig:crossvalidation-1}a-c are not meaningful, only
differences matter. 

CG MD simulations are generated for the selected models by iterating
Eq. (\ref{eq:Smoluchowski}). For each model, one hundred independent
simulations starting from structures sampled randomly from the atomistic
simulation are performed for 1 million steps each, and the aggregated
data are used to produce the free energy as a function of the dihedral
coordinates. Fig. \ref{fig:Free-energy} compares the free energy
computed via (\ref{eq:free_energy_data}) from the underlying atomistic
MD simulations and the free energy resulting from the selected CG models. 
Only the regularized CGnet model can correctly reproduce the position
of the all main free energy minima (Fig. \ref{fig:Free-energy}a,
c). On the contrary, the spline model is not able to capture the shallow
minima corresponding to positive values of the dihedral angle $\phi$,
and introduces several spurious minima (Fig. \ref{fig:Free-energy}b).
This comparison confirms that selecting CG models by minimal mean
force prediction error achieves models that are better from a physical
viewpoint. 

As an \textit{a posteriori} analysis of the results 
we have performed MD simulation for the CG models corresponding to
different choices of hyper-parameters, both for the spline model and
CGnet. For each choice of hyper-parameters, we have computed the difference
between the free energy as a function of the dihedral angles resulting
from the CG simulations and the one from the all-atom models. Differences
in free energy were estimated by discretizing the space spanned by
the two dihedral angles and computing the mean square difference on
all bins. The difference between a given model and CGnet was computed
by shifting the free energy of CGnet by a constant value that minimizes
the overall mean square difference. The free energy difference for
the spline models is always significantly larger than for the CGnet
models (Fig. \ref{fig:crossvalidation-1}d-f). Interestingly, the
minima in the difference in free energy correspond to the minima in
the cross-validation curves reported in Fig. \ref{fig:crossvalidation-1}a-c,
and the optimal values of hyper-parameters selected by cross-validation
yield the absolute minimum in the free energy difference (indicated
in Fig. \ref{fig:crossvalidation-1}f as C5 for CGnet and S4 for
the spline model).
This point is illustrated more explicitly in the SI (Section E, Figs. S4, S5), and
demonstrates that the cross-validation error of different models are correlated
with errors in approximating the two-dimensional free energy surface of alanine
dipeptide.

For the CGnet, regularization is extremely important: without regularization
the free energy only matches near the most pronounced minima and unphysical
structures are sampled outside (Fig. \ref{fig:Free-energy}d and SI
Section D). With regularization, these unphysical regimes are avoided,
all sampled structures appear chemically valid (Fig. \ref{fig:Free-energy}e)
and the distributions of bonds and angles follow those in the atomistic
simulations (SI Section D, Fig. S3).

\subsection{Coarse-graining of Chignolin folding/unfolding in water}

\begin{figure}[!h]
\begin{centering}
\includegraphics[width=1\columnwidth]{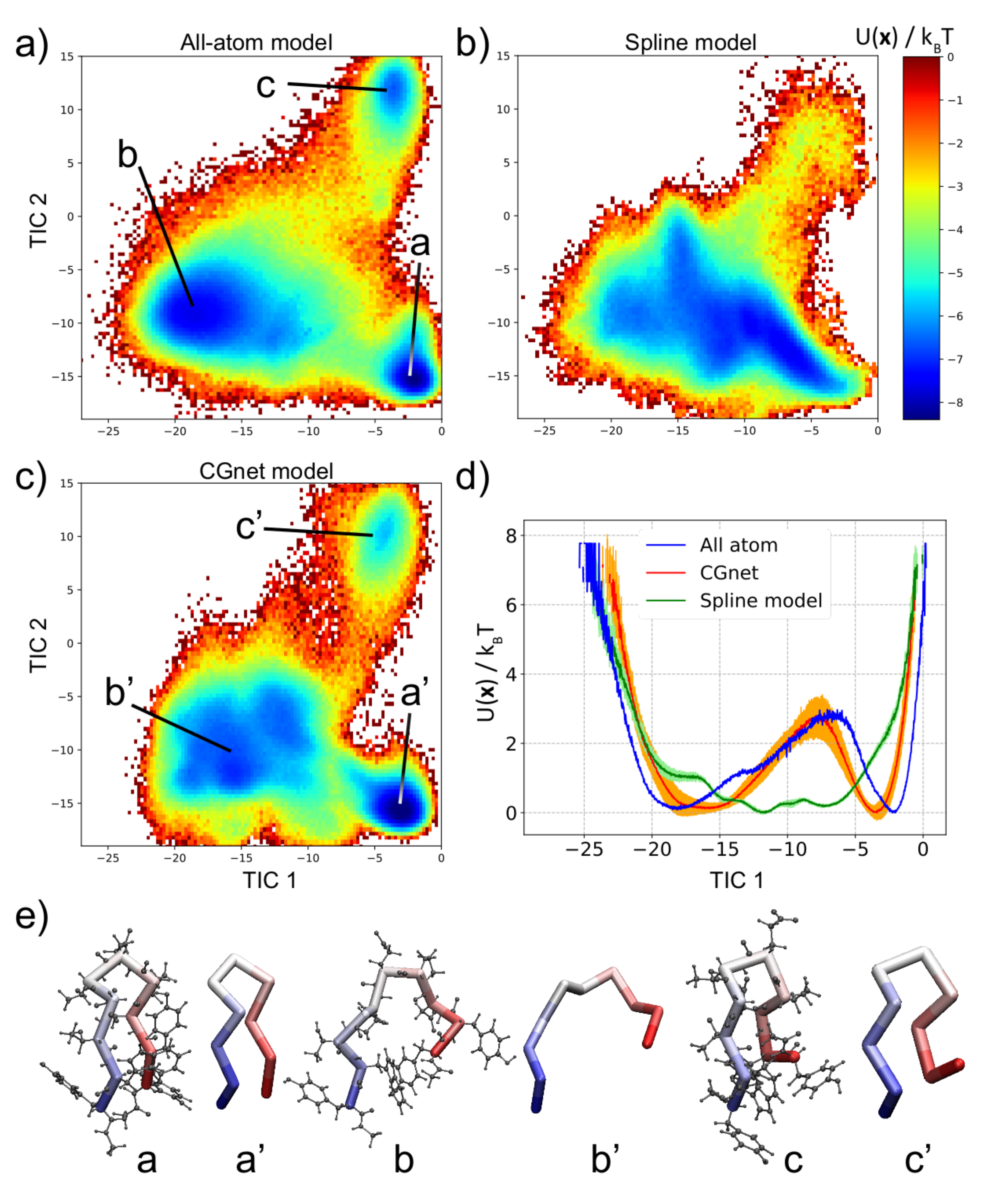}
\par\end{centering}
\caption{\label{fig:Free-energy-Chignolin}
\footnotesize
Free energy landscape of Chignolin
for the different models. (a) The free energy as obtained from all-atom
simulation, as a function of the first two TICA coordinates. (b) The
free energy as obtained from the spline model, as a function of the
same two coordinates used in the all-atom model. (c) The free energy
as obtained from CGnet, as a function of the same two coordinates.
(d) Comparison of the one dimensional free energy as a function of
the first TICA coordinate (corresponding to the folding/unfolding
transition) for the three models: all-atom (blue), spline (green),
and CGnet (red). (e) Representative Chignolin configurations in the
three minima from all-atom simulation (a, b, c) and CGnet (a', b',
c').}
\end{figure}

Finally, we test the CGnet on a much more challenging problem: the
folding/unfolding dynamics of the polypeptide Chignolin in water.
Chignolin consists of 10 amino acids plus termini and exhibits a clear
folding/unfolding transition. The all-atom model contains 1881 water
molecules, salt ions and the Chignolin molecule, resulting in nearly
6000 atoms. To focus on the folding/unfolding transition, data was
generated at the melting temperature 350 K, mimicking the setup used
for the Anton supercomputer simulation in \citep{Lindorff-Larsen2011}.
To obtain a well-converged ground truth, we generated 3742 short MD
simulations with an aggregate length of 187.2 $\mu$s on GPUgrid \citep{Buch2010}
using the ACEMD program \citep{Harvey2009}. The free energy landscape
is computed on the two collective variables describing the slowest
processes, computed by the TICA method \citep{PerezEtAl_JCP13_TICA}.
Since the individual MD simulations are too short to reach equilibrium,
the equilibrium distribution was recovered by reweighting all data
using a Markov state model \citep{PrinzEtAl_JCP10_MSM1}. See SI for
details on the MD simulation and Markov model analysis.

Fig. \ref{fig:Free-energy-Chignolin}a shows the free energy as a
function of the first two TICA coordinates. Three minima are clearly
identifiable on this free energy landscape: states a (folded), b (unfolded)
and c (partially misfolded), ordered alphabetically from most to least
populated. Representative configurations in these minima as shown
in Fig. \ref{fig:Free-energy-Chignolin}e. As a result, the first
TICA mode is a folding-unfolding coordinate, while the second is a
misfolding coordinate.

Using a regularized CGnet, we coarse-grain the 6000-atom system to
$10$ CG beads representing the $\alpha$-carbons of Chignolin. Thus,
not only the polypeptide is coarse-grained, but also the solvation
free energy is implicitly included in the CG model. Similar to what
was done for alanine dipeptide, roto-translational invariance of the
energy was implemented by using a CGnet featurization layer that maps
the $C_{\alpha}$ Cartesian coordinates to all pairwise distances
between CG beads, all angles defined by three adjacent CG beads, and
the $cos$ and $sin$ of all the dihedral angles defined by four CG
adjacent beads. The regularizing baseline energy includes a harmonic
term for each bond and angle, and an excluded volume term for each
pairwise distance between CG particles that are separated by more
than two bonds.

Similar to the case of alanine dipeptide, a classical few-body spline model was
defined for comparison whose CG potential is a sum of bonded and non-bonded
terms, where each term is a nonlinear function of a single feature
(pairwise distances, angles, dihedrals).

Both CGnet and spline model are optimized through a five-stage cross-validation
search for the hyper-parameters, in the following order:
Depth $D$, Width $W$, exponent of the excluded volume term $c$,
radius of the excluded volume term $\sigma$, and Lipschitz regularization
strength $\lambda.$ The results of the cross-validation are shown
in Fig. S8. This optimization
resulted in the hyper-parameter values $D=5$, $W=250$, $c=6$, $\sigma=5.5$
and $\lambda=4.0$. For the spline model, the optimal values of the
hyper-parameters are $D=3$, $W=12$ (for each feature), $c=10$,
$\sigma=4.0$, and $\lambda=5.0$ (Fig. S8).
The potential resulting from CGnet and the spline model is then used
to run long simulations with Eq. (\ref{eq:Smoluchowski}). One hundred
simulations of 1 million steps each were generated using randomly
sampled configurations from the training data as starting points.
For comparison, the aggregated data are projected on the TICA coordinates
obtained from all-atom simulations and free energy landscapes are
computed directly using Eq. (\ref{eq:free_energy_data}) (Fig. \ref{fig:Free-energy-Chignolin}b,
c). For a more quantitative comparison, the free energies are also
reported along the first TICA coordinate that indicates folding /
unfolding (Fig. \ref{fig:Free-energy-Chignolin}d).

These figures clearly show that the spline model cannot reproduce
the folding/unfolding dynamics of Chignolin, as the folded and unfolded
states are not well defined (Fig. \ref{fig:Free-energy-Chignolin}b,
d). On the contrary, CGnet not only can consistently fold and unfold
the protein but also correctly identifies three well defined minima:
the folded (a'), unfolded (b'), and partially misfolded (c') ensembles
corresponding to the minima a, b, and c in the all-atom fully solvated
model (Fig. \ref{fig:Free-energy-Chignolin}c,d). Representative structures
in the three minima are shown in Fig. \ref{fig:Free-energy-Chignolin}e:
the structures obtained from the CGnet simulations are remarkably
similar to the ones obtained in the all-atom simulations. These results
reinforce what has been already observed for alanine dipeptide above: the multi-body
interactions captured by CGnet are essentialy for correct reproduction of the
free energy landscape for the protein Chignolin.
The absence of
such interactions in the spline model dramatically alters the corresponding
free energy landscape to the point that the model can not reproduce
the folding/unfolding behavior of the protein.

\section{Conclusions}

Here we have formulated coarse-graining based on the force-matching
principle as a machine learning method. An important consequence of
this formulation is that coarse-graining is a supervised learning
problem whose loss function can be decomposed into the standard terms
of statistical estimator theory: Bias, Variance and Noise. These terms
have well-defined physical meanings and can be used in conjunction
with cross-validation in order to select model hyper-parameters and
rank the quality of different coarse-graining models.

We have also introduced CGnets, a class of neural networks that can
be trained with the force matching principle and can encode all physically
relevant invariances and constraints: (1) invariance of the free energy
and mean force with respect to translation of the molecule, (2) invariance
of the free energy and equivariance of the mean force with respect
to rotation of the molecule, (3) 
the mean force is a conservative
force field generated by the free energy, and (4) a prior energy can
be applied in order to prevent the simulations with CGnets to diverge
into unphysical state space regions outside the training data, such
as states with overstretched bonds or clashing atoms. Future CGnets
may include additional invariances, such as permutational invariance
of identical CG particles, e.g. permutation of identical particles
in symmetric rings.

The results presented above show that CGnet can be used to define
effective energies for CG models that optimally reproduce the equilibrium
distribution of a target atomistic model. CGnet provides a better
approximation than functional forms commonly used for CG models as
it automatically includes multi-body effects and non-linearities.
The work presented here provides a proof of principle for this approach
on relatively small solutes, but already demonstrates that the complex
solvation free energy involved in the folding/unfolding of a polypeptide
such as Chignolin can be encoded in a CGnet consisting of only the
$C_{\alpha}$ atoms. The extension to larger and more complex molecules
presents additional challenges and may require to include additional
terms to enforce physical constraints.

Additionally, the CG model considered here is designed \textit{ad
hoc} for a specific molecule and is not transferable to the study
of different systems. Transferability remains an outstanding issue
in the design of coarse grained models \citep{Noid2013} and its requirement
may decrease the ability to reproduce faithfully properties of specific
systems \citep{Johnson2007,Mullinax2009b,Thorpe2011,Wang2009,Allen2009}.
In principle, transferable potentials can be obtained by designing
input features for CGnet imposing a dependence of the energy function
on the CG particle types and their environment \citep{Mullinax2009b},
similarly to what is done in the learning of potential energy functions
from quantum mechanics data (see e.g. \citep{BehlerParrinello_PRL07_NeuralNetwork,BartokKondorCsanyi_PRB13_SOAP,Smith2017,Schuett2017,ZhangHan2018}).
This approach may be able to define transferable functions if enough
data are used in the training \citep{Schuett2017,ZhangHan2018}. We
leave the investigation on the trade-off between transferability and
accuracy for future studies.

It is also important to note that the formulation used here to define
an optimal CG potential aims at reproducing structural properties
of the system, but it does not determine the equations for its dynamical
evolution. If one is interested in designing CG models that can reproduce
molecular dynamical mechanisms, e.g. to reproduce the slow dynamical
processes of the fine-grained model, alternative approaches need to
be investigated.

\begin{acknowledgement}
We thank Alex Kluber, Justin Chen, Lorenzo Boninsegna, Eugen Hruska,
and Feliks N\"uske for comments on the manuscript. This work was supported
by the National Science Foundation (CHE-1265929, CHE-1738990, and
PHY-1427654), the Welch Foundation (C-1570), the MATH+ excellence
cluster (AA1-6, EF1-2), the Deutsche Forschungsgemeinschaft (SFB 1114/C03, SFB
958/A04, TRR 186/A12),
the European Commission (ERC CoG 772230 ``ScaleCell''), the Einstein
Foundation Berlin (Einstein Visiting Fellowship to CC), and the Alexander von
Humboldt foundation (Postdoctoral fellowship to SO). Simulations
have been performed on the computer clusters of the Center for Research
Computing at Rice University, supported in part by the Big-Data Private-Cloud
Research Cyberinfrastructure MRI-award (NSF grant CNS-1338099), and
on the clusters of the Department of Mathematics and Computer Science
at Freie Universit\"at, Berlin.
GDF acknowledges support from MINECO (Unidad de Excelencia Mar\'{i}a de
Maeztu MDM-2014-0370 and BIO2017-82628-P) and FEDER. This project has received
funding from the European Union's Horizon 2020 research and innovation
programme under grant agreement No 675451 (CompBioMed project).
We thank the GPUGRID donors for their compute time.
\end{acknowledgement}

\begin{suppinfo}
Supporting Information is available providing:
\begin{itemize}
\item
A derivation of equation (\ref{eq:force-matching-decomp1}) 
\item
Cross-validation for all the models presented in the manuscript
\item
Additional details on the training of the models
\item
Distribution of bonds and angles for the different models of alanine dipeptide
\item
Changes in the free energy of alanine dipeptide with different hyper-parameters
\item
Energy decomposition for the CGnet model of alanine dipeptide
\item
Details on Chignolin setup and simulation
\item 
Markov State Model analysis of Chignolin all-atom simulations
\end{itemize}
\end{suppinfo}


\providecommand{\latin}[1]{#1}
\makeatletter
\providecommand{\doi}
  {\begingroup\let\do\@makeother\dospecials
  \catcode`\{=1 \catcode`\}=2 \doi@aux}
\providecommand{\doi@aux}[1]{\endgroup\texttt{#1}}
\makeatother
\providecommand*\mcitethebibliography{\thebibliography}
\csname @ifundefined\endcsname{endmcitethebibliography}
  {\let\endmcitethebibliography\endthebibliography}{}




\clearpage
\newpage

\maketitle
\onecolumn

\setcounter{page}{1}
\setcounter{table}{0}
\setcounter{figure}{0}
\renewcommand{\thepage}{S\arabic{page}}   
\renewcommand{\thetable}{S\arabic{table}}    
\renewcommand{\thefigure}{S\arabic{figure}}

\section*{Supplementary Material}

\subsection{Decomposition of the force matching error}

The decomposition of the force matching error (\ref{eq:force-matching})
can be achieved by adding and subtracting the mean force (\ref{eq:mean_force})
and splitting the norm: 
\begin{eqnarray*}
\chi^{2}\left[U(\mathbf{x})\right] & = & \left\langle \left\langle \left\Vert \xi(\mathbf{F}(\mathbf{r}))-\mathbf{f}(\mathbf{x})+\mathbf{f}(\mathbf{x})+\nabla U(\mathbf{x})\right\Vert ^{2}\right\rangle _{\mathbf{r}\mid\mathbf{x}}\right\rangle _{\mathbf{x}}\\
 & = & \left\langle \left\langle \left\Vert \xi(\mathbf{F}(\mathbf{r}))-\mathbf{f}(\mathbf{x})\right\Vert ^{2}\right\rangle _{\mathbf{r}\mid\mathbf{x}}\right\rangle _{\mathbf{x}}+\left\langle \left\Vert \mathbf{f}(\mathbf{x})+\nabla U(\mathbf{x})\right\Vert ^{2}\right\rangle _{\mathbf{x}}\\
 & \: & +2\left\langle \left\langle (\xi(\mathbf{F}(\mathbf{r}))-\mathbf{f}(\mathbf{x}))^{\top}(\mathbf{f}(\mathbf{x})+\nabla U(\mathbf{x}))\right\rangle _{\mathbf{r}\mid\mathbf{x}}\right\rangle _{\mathbf{x}}.
\end{eqnarray*}

This expression is equivalent to Eq. (\ref{eq:force-matching-decomp1}).
as the mixed term is zero:
\begin{align*}
 & \left\langle \left\langle (\xi(\mathbf{F}(\mathbf{r}))-\mathbf{f}(\mathbf{x}))^{\top}(\mathbf{f}(\mathbf{x})+\nabla U(\mathbf{x}))\right\rangle _{\mathbf{r}\mid\mathbf{x}}\right\rangle _{\mathbf{x}}\\
 & =\left\langle \mathbf{f}(\mathbf{x})^{\top}\mathbf{f}(\mathbf{x})\right\rangle _{\mathbf{x}}+\left\langle \mathbf{f}(\mathbf{x})^{\top}\nabla U(\mathbf{x})\right\rangle _{\mathbf{x}}\\
 & \:\:\:\:-\left\langle \mathbf{f}(\mathbf{x})^{\top}\mathbf{f}(\mathbf{x})\right\rangle _{\mathbf{x}}-\left\langle \mathbf{f}(\mathbf{x})^{\top}\nabla U(\mathbf{x})\right\rangle _{\mathbf{x}}\\
 & =0
\end{align*}

The decomposition of the expected prediction error in the form of
Eq. (\ref{eq:bias_variance_noise}) can be achieved by adding and subtracting
the mean estimator $\bar{\mathbf{f}}(\mathbf{\mathbf{X}})=\mathbb{E}\left[-\nabla U(\mathbf{X};\boldsymbol{\theta})\right]$:

\begin{align*}
\mathbb{E}\left[L(\boldsymbol{\theta};\mathbf{R})\right] & =\mathbb{E}_{\mathbf{R}\mid\mathbf{X}}\left[\left\Vert \mathbf{f}(\mathbf{X})+\nabla U(\mathbf{X};\boldsymbol{\theta})\right\Vert _{F}^{2}\right]+\mathrm{Noise}\\
 & =\mathbb{E}\left[\left\Vert
 \underset{A}{\underbrace{\left(\mathbf{f}(\mathbf{X})-\bar{\mathbf{f}}(\mathbf{\mathbf{X}})\right)}}+\underset{B}{\underbrace{\left(\bar{\mathbf{f}}(\mathbf{\mathbf{X}})+\nabla U(\mathbf{X};\boldsymbol{\theta})\right)}}\right\Vert _{F}^{2}\right]+\mathrm{Noise}\\
 & =\mathbb{E}\left[\left\Vert A\right\Vert _{F}^{2}\right]+\mathbb{E}\left[\left\Vert B\right\Vert _{F}^{2}\right]+2\mathbb{E}\left[
 \sum_{i,j}\left(A*B \right)_{i,j}
 \right]+\mathrm{Noise},
\end{align*}
where $*$ is the element-wise product. We follow standard results
for regression. For the mixed term we can use 
\[
\mathbb{E}\left[\sum_{i,j}\left(A*B \right)_{i,j} \right] =
\sum_{i,j} \mathbb{E}\left[\left(A*B\right)_{i,j}\right] = \sum_{i,j} \left( \mathbb{E}\left[A*B \right] \right)_{i,j}
\]
and this expectation value disappears:
\begin{align*}
\mathbb{E}\left[A*B\right] &
=\mathbb{E}\left[\left(\mathbf{f}(\mathbf{X})-\bar{\mathbf{f}}(\mathbf{\mathbf{X}})\right)*\left(\bar{\mathbf{f}}(\mathbf{\mathbf{X}})+\nabla U(\mathbf{X};\boldsymbol{\theta})\right)\right]\\
 & =\mathbb{E}\left[\mathbf{f}(\mathbf{X})\right]*\bar{\mathbf{f}}(\mathbf{\mathbf{X}})+\mathbb{E}\left[\mathbf{f}(\mathbf{X})*\nabla
 U(\mathbf{X};\boldsymbol{\theta})\right]-\mathbb{E}\left[\bar{\mathbf{f}}(\mathbf{\mathbf{X}})*\bar{\mathbf{f}}(\mathbf{\mathbf{X}})\right]-\bar{\mathbf{f}}(\mathbf{\mathbf{X}})*\mathbb{E}\left[\nabla U(\mathbf{X};\boldsymbol{\theta})\right]\\
 &
 =\mathbf{f}(\mathbf{X})*\bar{\mathbf{f}}(\mathbf{\mathbf{X}})-\mathbf{f}(\mathbf{X})*\bar{\mathbf{f}}(\mathbf{\mathbf{X}})-\bar{\mathbf{f}}(\mathbf{\mathbf{X}})*\bar{\mathbf{f}}(\mathbf{\mathbf{X}})+\bar{\mathbf{f}}(\mathbf{\mathbf{X}})*\bar{\mathbf{f}}(\mathbf{\mathbf{X}})\\
 & =0.
\end{align*}
The remaining terms define bias and variance.

\subsection{Cross-validation for the coarse-graining of the 2d toy model}

We report here the results from cross-validation for the choice of
hyper-parameters for the coarse-graining of the 2d toy model discussed in the main text.

\begin{figure}[h]
\begin{centering}
\includegraphics[width=0.5\columnwidth]{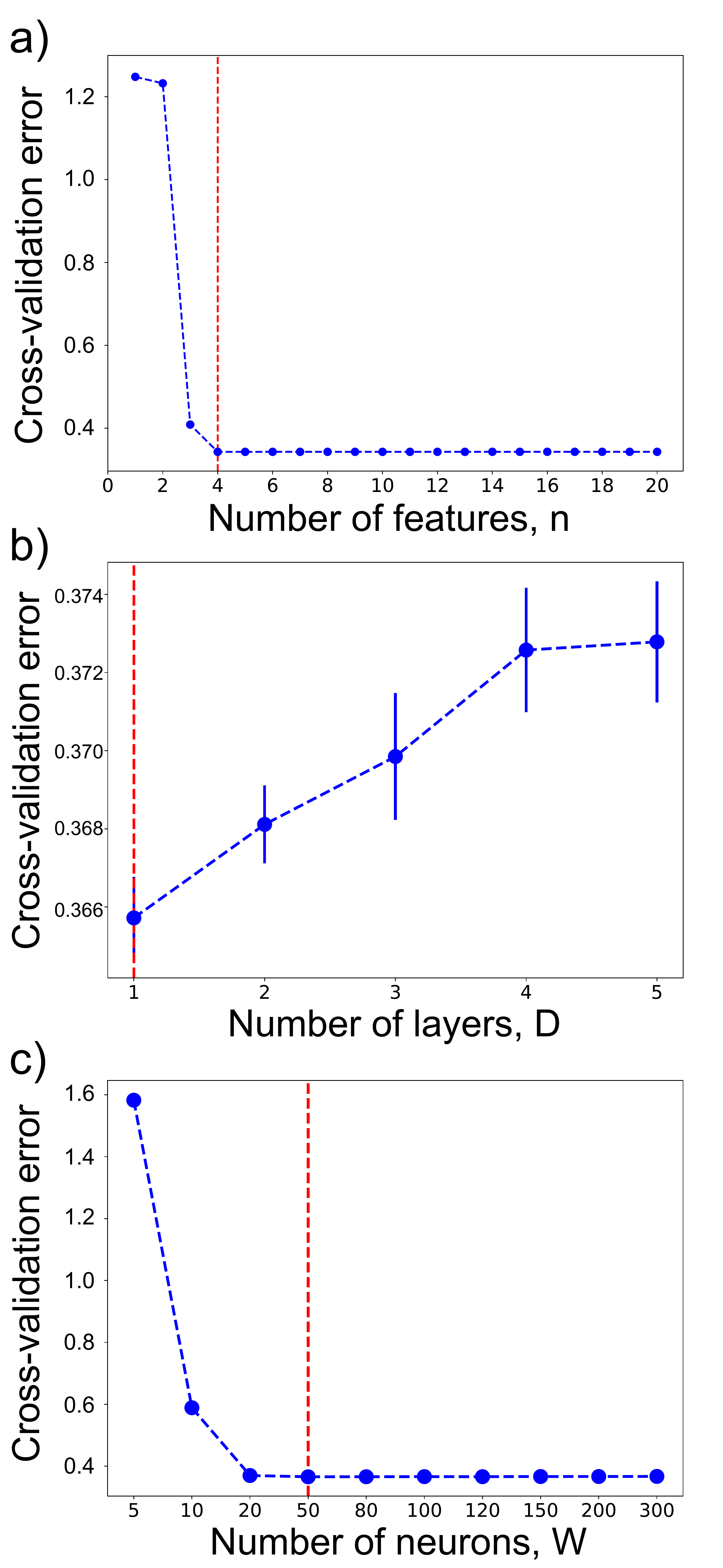}
\par\end{centering}
\caption{\label{fig:2DCV}
\footnotesize Model selection for CG model of 2D system using cross-validation.
a) Choice of the set of feature functions for feature regression.
b) First stage of regularized CGnet hyper-parameter selection: the
optimal number of layers, $D$. c). Second stage of regularized CGnet
hyperparameter selection: the optimal number of neurons per layer,
$W$. Red dashed lines indicate the minimal cross-validation error.
Error bars represent the standard error of the mean cross-validation
error over five cross-validation folds, in panels a) and c) the error
bars are invisible as they are smaller than the marker. The unit of
the cross-validation error is $(k_{B}T)^{2}$ , with the unit of length
equal to 1.}
\end{figure}

The feature regression for the coarse-graining of the 2 dimensional
toy model is performed with the twenty basis functions listed in Table
\ref{tab:2d-features} selected as features. Cross-validation is performed
with the Stepwise Sparse Regressor introduced in \citep{Boninsegna2018}.
The minimum cross-validation error is obtained when the first four
functions are used as features, as shown in Fig. \ref{fig:2DCV}.

\begin{table}[h]
\caption{\label{tab:2d-features}Twenty elementary basis functions.}

\centering{}%
\begin{tabular}{|c|c||c|c|}
\hline 
function ID & function, $f(x)$ & function ID & function, $f(x)$\tabularnewline
\hline 
\hline 
1 & $1$ & 11 & $x^{10}$\tabularnewline
\hline 
2 & $x$ & 12 & $sin(x)$\tabularnewline
\hline 
3 & $x^{2}$ & 13 & $cos(x)$\tabularnewline
\hline 
4 & $x^{3}$ & 14 & $sin(6x)$\tabularnewline
\hline 
5 & $x^{4}$ & 15 & $cos(6x)$\tabularnewline
\hline 
6 & $x^{5}$ & 16 & $sin(11x)$\tabularnewline
\hline 
7 & $x^{6}$ & 17 & $cos(11x)$\tabularnewline
\hline 
8 & $x^{7}$ & 18 & $tanh(10x)$\tabularnewline
\hline 
9 & $x^{8}$ & 19 & $tanh^{2}(10x)$\tabularnewline
\hline 
10 & $x^{9}$ & 20 & $e^{-50x^{2}}$\tabularnewline
\hline 
\end{tabular}
\end{table}

The results from the cross-validation of the CGnet for the toy 2 dimensional system are reported
in Tables \ref{tab:hyper_2d} and Fig. \ref{fig:2DCV}.

\begin{table}[h]
\caption{\label{tab:hyper_2d}Hyper-parameter optimization for unregularized
CGnet of two-dimensional model system. $D$: network depth, $W:$
network width. The unit of the cross-validation error is $(k_{B}T)^{2}$
, with the unit of length equal to 1.}

\centering{}%
\begin{tabular}{|c|c|}
\hline 
$D$ ($W=20$) & Cross-validation error\tabularnewline
\hline 
\hline 
1 & \textbf{0.3785} $\pm$ \textbf{0.0024}\tabularnewline
\hline 
2 & 0.5457 $\pm$ 0.0973\tabularnewline
\hline 
3 & 0.7339 $\pm$ 0.0298\tabularnewline
\hline 
4 & 0.5695 $\pm$ 0.0172\tabularnewline
\hline 
5 & 0.8543 $\pm$ 0.1227\tabularnewline
\hline 
\multicolumn{1}{c}{} & \multicolumn{1}{c}{}\tabularnewline
\multicolumn{1}{c}{} & \multicolumn{1}{c}{}\tabularnewline
\multicolumn{1}{c}{} & \multicolumn{1}{c}{}\tabularnewline
\multicolumn{1}{c}{} & \multicolumn{1}{c}{}\tabularnewline
\end{tabular} %
\begin{tabular}{|c|c|}
\hline 
$W$ ($D=1$) & Cross-validation error\tabularnewline
\hline 
\hline 
5 & 0.5674 $\pm$ 0.0044\tabularnewline
\hline 
10 & 0.8762 $\pm$ 0.0048\tabularnewline
\hline 
20 & 0.3785 $\pm$ 0.0024\tabularnewline
\hline 
40 & 0.3729 $\pm$ 0.0017\tabularnewline
\hline 
60 & 0.3703 $\pm$ 0.0013\tabularnewline
\hline 
80 & 0.3682 $\pm$ 0.0013\tabularnewline
\hline 
100 & 0.3671 $\pm$ 0.0013\tabularnewline
\hline 
120 & \textbf{0.3661} $\pm$ \textbf{0.0012}\tabularnewline
\hline 
150 & 0.3661 $\pm$ 0.0012\tabularnewline
\hline 
\end{tabular}
\end{table}

\clearpage

\subsection{Training CG models}

Networks were optimized using the Adam adaptive stochastic gradient
descent method \citep{KingmaBa_ADAM} with default settings using
the PyTorch program. The batch-size was 128 for the 2D model and 512
for alanine dipeptide. The convergence of the training error and validation
error for the 2d toy model and alanine dipeptide is shown in Fig.
\ref{fig:Training-error} below.

\begin{figure}[h]
\begin{centering}
\includegraphics[width=0.5\columnwidth]{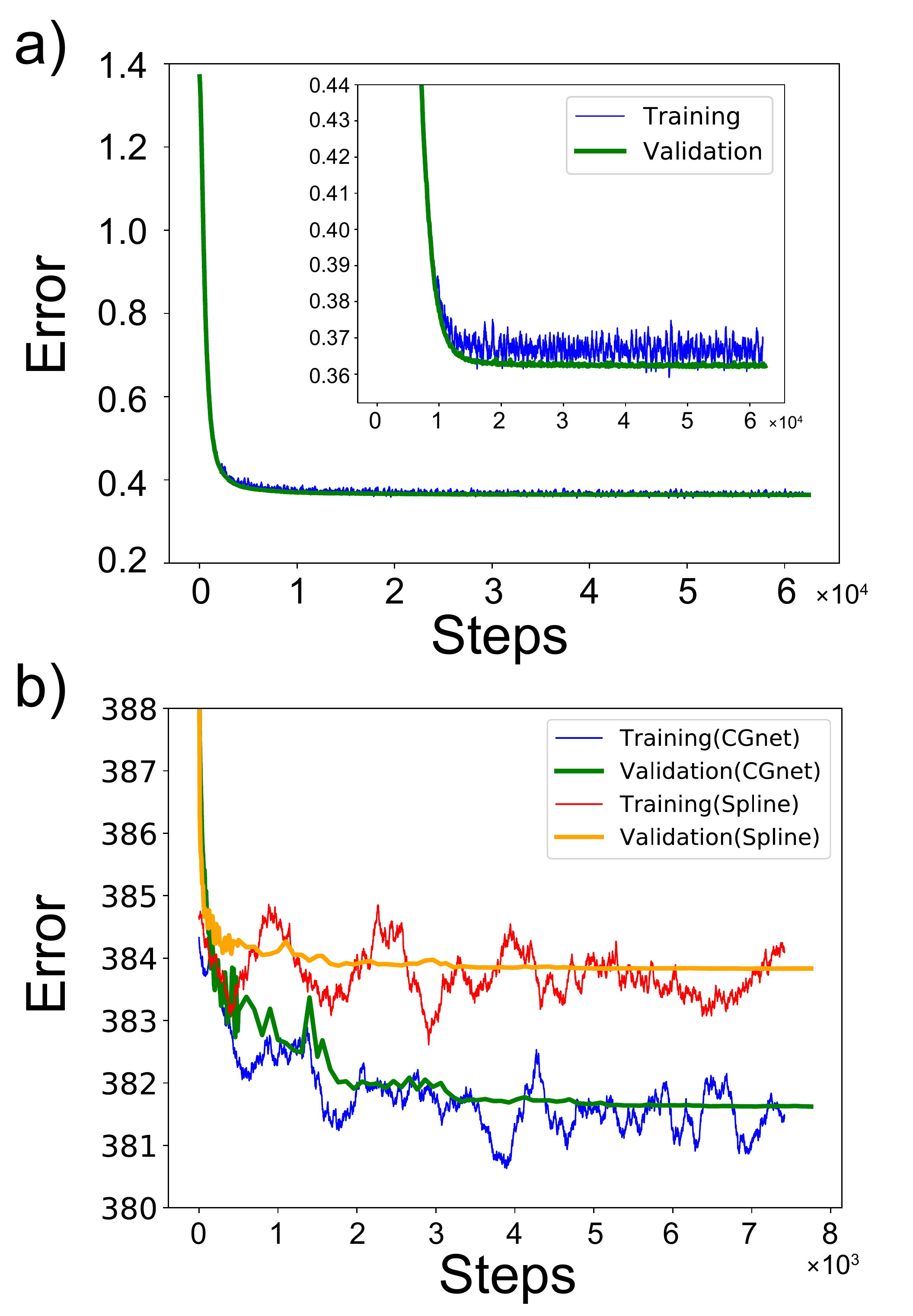}
\par\end{centering}
\caption{\label{fig:Training-error}
\footnotesize Training error and validation error for
(a) the 2D model and (b) alanine dipeptide. In (a), the model
is the regularized CGnet, in (b), the model is the regularized CGnet
and the spline model, which is also regularized. All errors are averaged
over 200000 points -- for the training error this is done by averaging
over the most recent batches, while the validation error is shown
for a fixed validation set. Note that the hyper-parameter choices
are made via cross-validation. The unit of the error is $(k_{B}T)^{2}$
in (a) and $[kcal/(mol\mathring{\cdot\textrm{A}})]^{2}$ in (b).}
\end{figure}

\clearpage

\subsection{Distribution of bond distances and angles for the different models
of alanine dipeptide}

\begin{figure}[h]
\begin{centering}
\includegraphics[width=0.95\columnwidth]{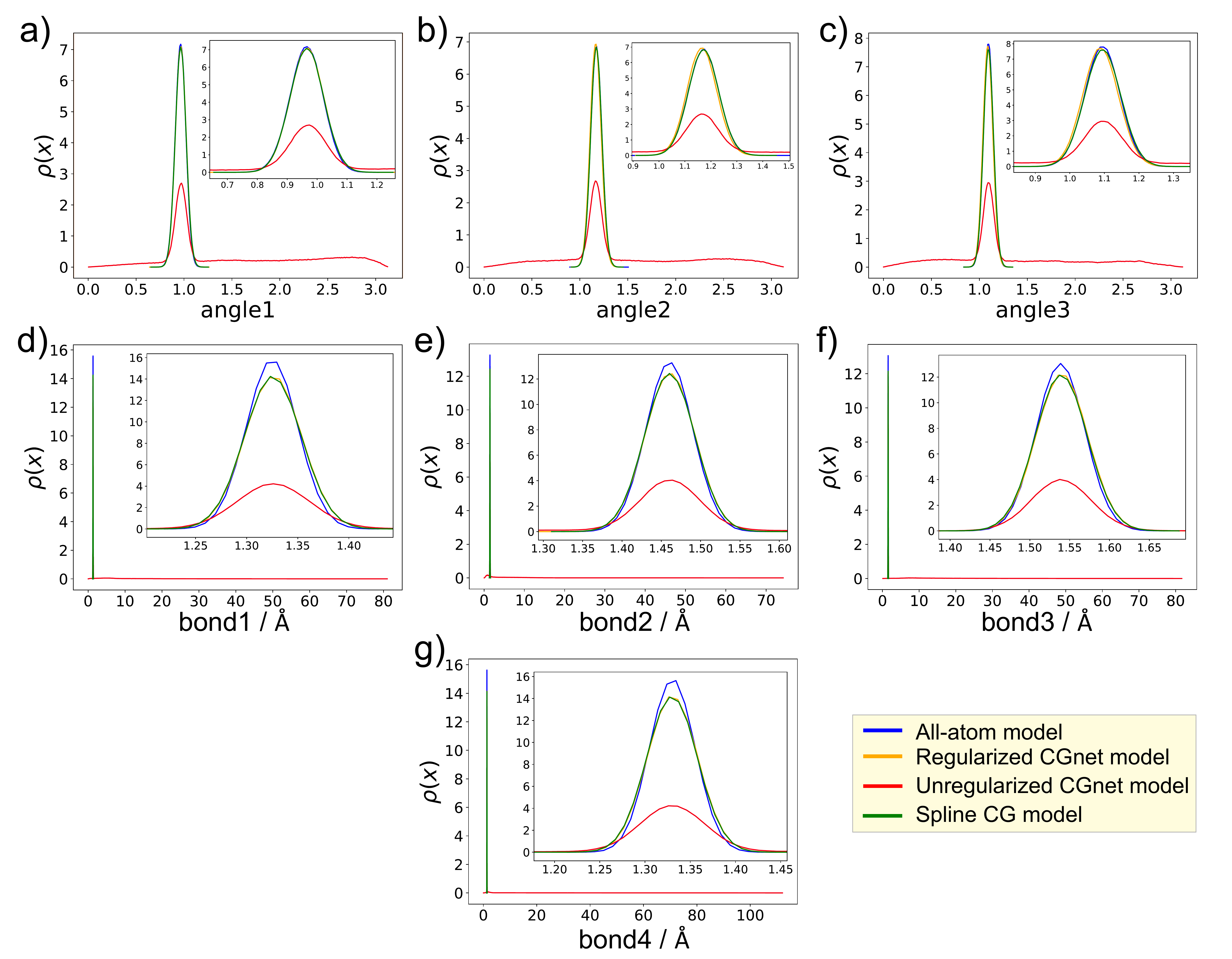}
\par\end{centering}
\caption{\label{fig:bonds-angles}
\footnotesize Probability density distribution for three
angles a)- c), and four bonds d)-g) for the alanine dipeptide models. Each panel contains the distribution
from four models: All-atom model (blue), regularized CGnet model (red),
unregularized CGnet model (cyan), spline CG model (green). The distribution
for regularized CGnet and spline model (with regularization) agree
with the true all-atom one. The distribution for the unregularized
CGnet has a wide range, which makes the distributions for the other
models appear very narrow in d)-g). The insets in d)-g) present zoomed
views of the distributions in the correct range.}
\end{figure}

\clearpage

\subsection{Changes in the free energy of alanine dipeptide with different hyper-parameters}

\begin{figure}[h]
\begin{centering}
\includegraphics[width=0.6\columnwidth]{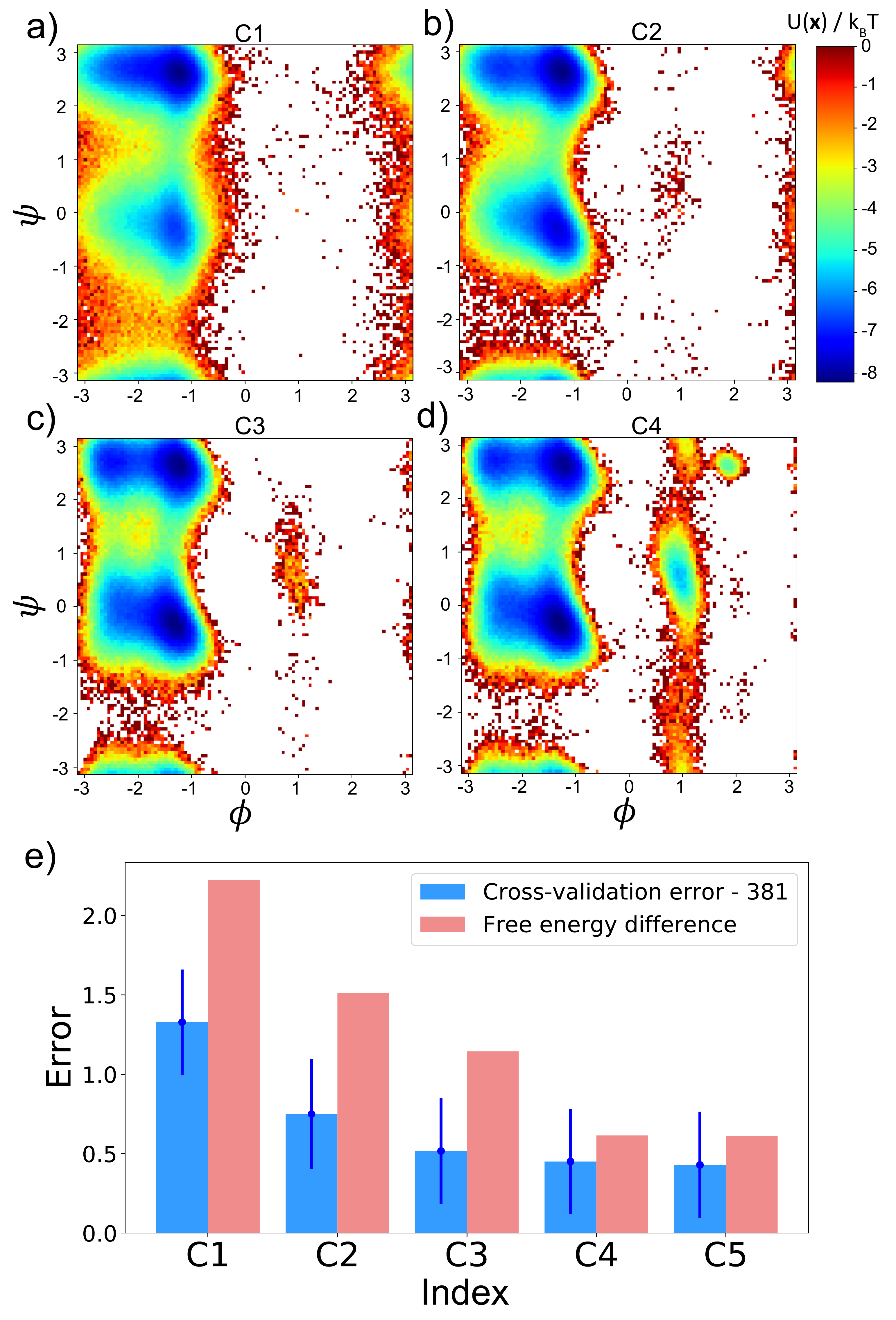}
\par\end{centering}
\caption{\label{fig:CGnet_FE_change-1}
\footnotesize Comparison of the free energy profiles
of CGnet models of alanine dipeptide with different choices of hyper-parameters. (a)-(d)
Free energy profiles with hyperparameters corresponding to the combination
indicated as C1, C2, C3, C4 in Fig. \ref{fig:crossvalidation-1}.
The choice of hyperparameters C5
correspond to the global optimum and is reported in Fig. \ref{fig:Free-energy}c.
(e) Comparison between the cross validation error (in $[kcal/(mol\mathring{\cdot\textrm{A}})]^{2}$)
and mean square free energy difference (in $[k_{B}T]^{2}$) for the
five selected hyperparameters. The value of 381 is subtracted from
the cross validation error to obtain values in the similar range as
the free energy differences.}
\end{figure}

\begin{figure}[h]
\begin{centering}
\includegraphics[width=0.6\columnwidth]{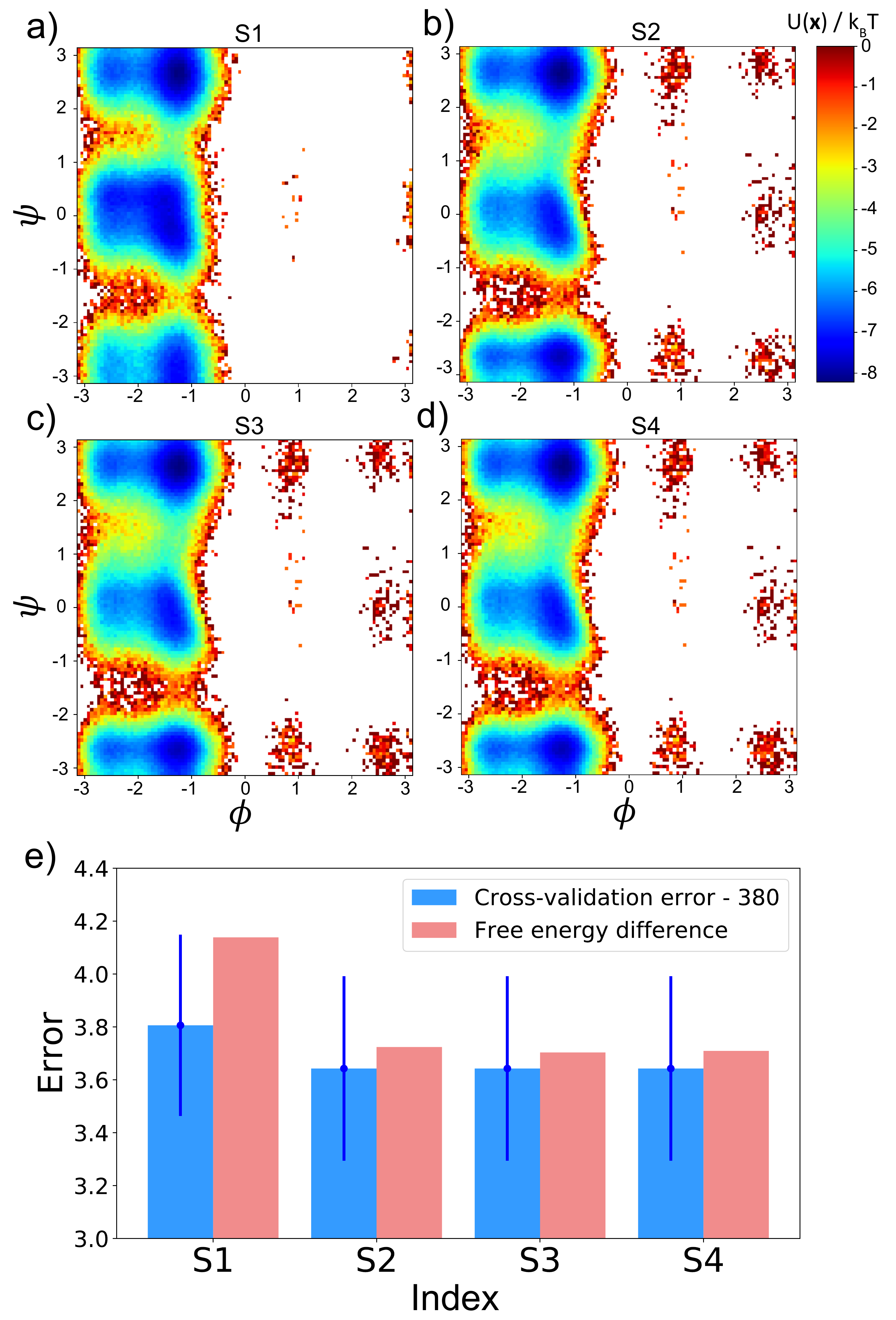}
\par\end{centering}
\caption{\label{fig:Spline_FE_change-1}
\footnotesize Comparison of the free energy profiles
of the spline models of alanine dipeptide with different choices of hyper-parameters. (a)-(d)
Free energy profiles with hyperparameters corresponding to the combination
indicated as S1, S2, S3, S4 in Fig. \ref{fig:crossvalidation-1}. 
The choice of hyperparameters S4
correspond to the global optimum and is also reported in Fig. \ref{fig:Free-energy}b.
(e) Comparison between the cross validation error (in $[kcal/(mol\mathring{\cdot\textrm{A}})]^{2}$)
and mean square free energy difference (in $[k_{B}T]^{2}$) for the
five selected hyperparameters. The value of 380 is subtracted from
the cross validation error to obtain values in the similar range as
the free energy differences.}
\end{figure}

In order to show how the free energy is approximating the atomistic free energy
as the hyper-parameters gradually reach the optimal values, we select
five hyper-parameters for CGnet (C1, C2, C3, C4, C5) and four for the
Spline model (S1, S2, S3, S4), as indicated in Fig. \ref{fig:crossvalidation-1}
in the manuscript.
For each of these combinations
of hyper-parameters, we report the corresponding two dimensional free
energy profiles in Fig. \ref{fig:CGnet_FE_change-1} and Fig. \ref{fig:Spline_FE_change-1}
(in addition to the free energy profile for the global optima reported
in Fig. \ref{fig:Free-energy}). The figures show that as the hyper-parameters
get closer to the optimal values the model free energy landscape becomes
closer to the atomistic free energy landscape.

\clearpage

\subsection{Energy decomposition for the CGnet model of alanine dipeptide.}

\begin{figure}[h]
\begin{centering}
\includegraphics[width=1\columnwidth]{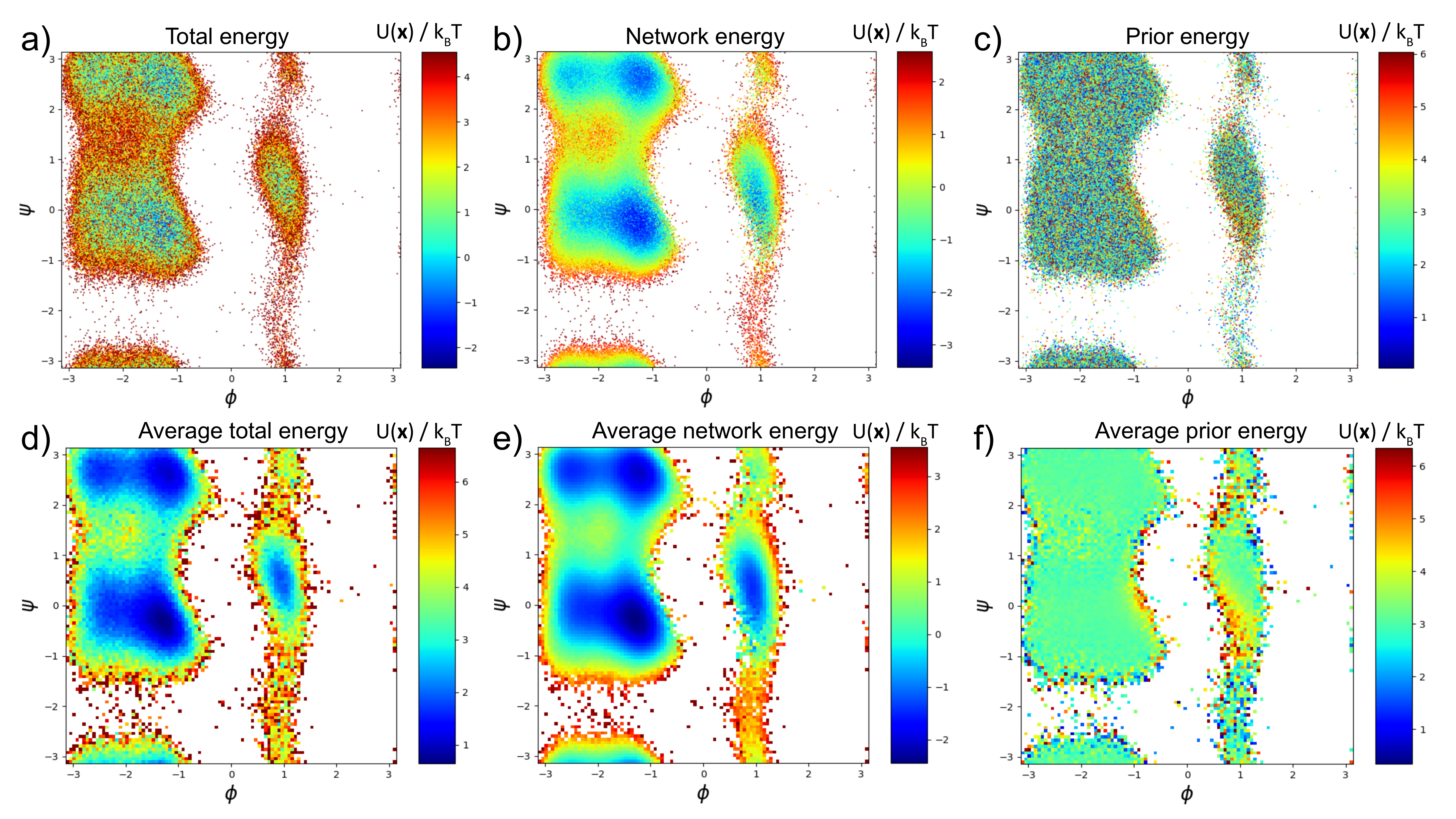}
\par\end{centering}
\caption{\label{fig:CGnet-energy-break}
\footnotesize 
CGnet energy decomposition for the
alanine dipeptide. In each simulated point, the total CGnet energy
(a) is decomposed in the energy contribution from the dense net (b),
and the baseline (or prior) energy (c). In each bin in the dihedral
angles space, the average total total energy (d) is decomposed into
the average dense net energy (e), and average prior energy (f).}
\end{figure}

As discussed in the main text, the use of a baseline energy to enforce
physical constraints plays an important role in the CGnet model. Here
we report the decomposition of the total CGnet energy into the contribution
of the baseline (prior) energy and the energy of the neural network. Figs.
\ref{fig:CGnet-energy-break}a-c report the decomposition for each
point sampled in the simulations performed with CGnet. Fig. \ref{fig:CGnet-energy-break}d-f
report the same quantity averaged over different bins in the space
spanned by the dihedral angles. The figures show that the network
energy captures the overall features of the free energy landscape
for this molecule, while the prior energy seems to play an important
role to enforce physical constraints mostly at the edges of the populated
regions in the landscape. This is in agreement with the intuition
that the prior energy term makes the system avoid high energy regions
not visited in the training data.

\subsection{Chignolin setup and simulation}

The initial structure of Chignolin was generated starting from the
cln025 peptide\citep{Honda2008}, with sequence TYR-TYR-ASP-PRO-GLU-THR-GLY-THR-TRP-TYR.
The structure was solvated in a cubic box of 40 $\text{Å}$, containing
1881 water molecules and two $\mathrm{Na}^{+}$ ions to neutralize
the peptide's negative charge, as described in \citep{Lindorff-Larsen2011}.
MD simulations were performed with ACEMD \citep{Harvey2009}, using
CHARMM22{*} \citep{Piana2011} force field and TIP3P \citep{Jorgensen1983}
water model at 350K temperature. A Langevin integrator was used with
a damping constant of 0.1 $\mathrm{ps}^{-1}$. Integration time step
was set to 4 fs, with heavy hydrogen atoms (scaled up to four times
the hydrogen mass) and holonomic constrains on all hydrogen-heavy
atom bond terms \citep{Feenstra1999}. Electrostatics were computed
using Particle Mesh Ewald with a cutoff distance of 9 $\text{Å}$
and grid spacing of 1 $\text{Å}$. Ten NVT simulations of 1 ns length
were carried out, with a dielectric constant of 80 and temperature
of 350K to generate ten different starting conformations for the production
runs. Production simulations consisted of 3744 independent simulations
of 50 ns, for a total aggregate time of 187.2 $\mu$s. All the simulations
were run using the GPUGRID \citep{Buch2010} distributed computing
platform. The first 1000 simulations were spawned from the 10 conformations
obtained previously. The remaining 2744 simulations were spawned using
the adaptive sampling \citep{Doerr2014} protocol implemented in HTMD
\citep{Doerr2016}. In adaptive sampling, multiple rounds of simulations
are performed, and each round the available trajectories are analyzed
to select the initial coordinates for the next round of simulations.
Each round was done every 10 to 20 simulations, respawning an equivalent
amount of new simulations. Initial coordinates for the respawned simulations
were selected proportionally to the inverse of the number of frames
per macrostate as explained in \citep{Doerr2016}. The Markov State
Model \citep{PrinzEtAl_JCP10_MSM1,SwopePiteraSuits_JPCB108_6571,ChoderaEtAl_JCP07,NoeHorenkeSchutteSmith_JCP07_Metastability,BucheteHummer_JPCB08}
constructed during the analysis was done using atom distances as projected
metric, TICA \citep{PerezEtAl_JCP13_TICA,SchwantesPande_JCTC13_TICA}
for dimensionality reduction method and k-Centers for clustering.
Force data used for training CGnet was obtained from the MD simulation
trajectories. ACEMD was used to read the Chignolin trajectories and
compute forces for all atoms for each simulation frame, using the
same parameters used for the MD simulations.

\subsection{Markov State Model analysis of Chignolin all-atom simulations}

\begin{figure}[h]
\begin{centering}
\includegraphics[width=0.5\columnwidth]{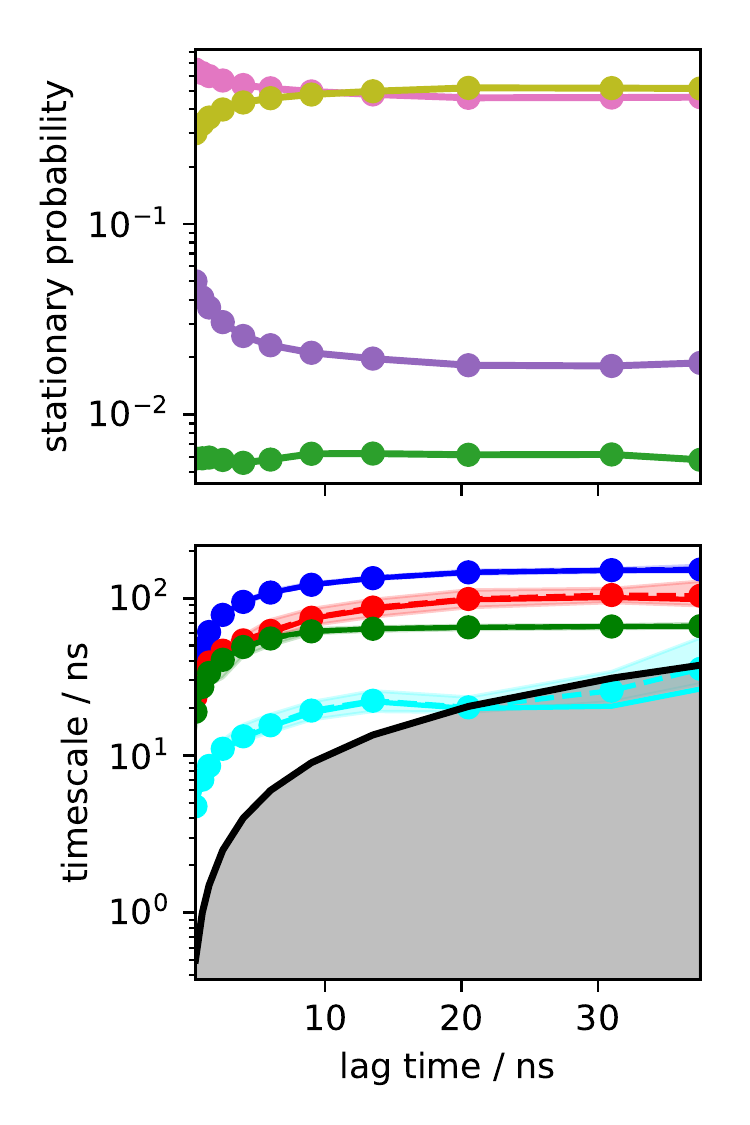}
\par\end{centering}
\caption{\label{fig:implied timescales}
Validation of a convergence of the
Chignolin all-atom Markov model, which is estimated at $\tau=37.5\,\mathrm{ns}$.
Top: Stationary probabilities of metastable states. Bottom: MSM implied
time scales.}
\end{figure}

MD simulation data of Chignolin from GPUGrid was featurized into all
pairwise C$_{\alpha}$ distances excluding pairs of nearest neighbors
residues (a total of 45 distances). Time-lagged independent component
analysis (TICA) \citep{PerezEtAl_JCP13_TICA,SchwantesPande_JCTC13_TICA}
was carried out with a lag $\tau=25\,\mathrm{ns}$. By using kinetic-map
\citep{NoeClementi_JCTC15_KineticMap,NoeClementi_JCTC16_KineticMap2}
and a kinetic variance cutoff of $95$\%, 4 TICs were retained for
further analysis. The 4 TICs were clustered into $350$ discrete states
using the $k$-means algorithm. All MD data was mapped onto their
discrete states and used for Markov state model (MSM) estimation.
The implied-timescales, $t_{i}=-\frac{\tau}{\log|\lambda_{i}|}$,
become constant as a function of lag-time ($\tau$) within statistical
uncertainty for lag-times above approximately $20\,\mathrm{ns}$.
Spectral analysis of a Markov state model estimated at a lagtime $\tau=37.5\,\mathrm{ns}$
reveal a spectral gap after the third implied-timescale suggesting
$4$ meta-stable states (Fig. \ref{fig:implied timescales}). Plotting
the populations of the meta-stable states as function of lag-time
show that these are stable for $\tau>10\,\mathrm{ns}$, and that three
of the four meta-stable states have significant probability mass $>1\%$.
These three most stable meta-stable states were used as reference
states a, b and c, ordered alphabetically from most to least populated
(shown in Fig. \ref{fig:Free-energy-Chignolin}a). To account for
the non-equilibrium nature of the multiple short molecular dynamics
trajectories, we used the estimated MSM ($\tau=37.5\,\mathrm{ns}$)
to reweighed data prior to calculating the reference free energy profiles.
These analyses were carried out using the PyEMMA \citep{SchererEtAl_JCTC15_EMMA2}
and MDTraj \citep{McGibbon_BJ15_MDTraj} software packages.

\clearpage

\subsection{Hyper-parameter optimization for Chignolin CG models}

\begin{figure}[h]
\begin{centering}
\includegraphics[width=0.9\columnwidth]{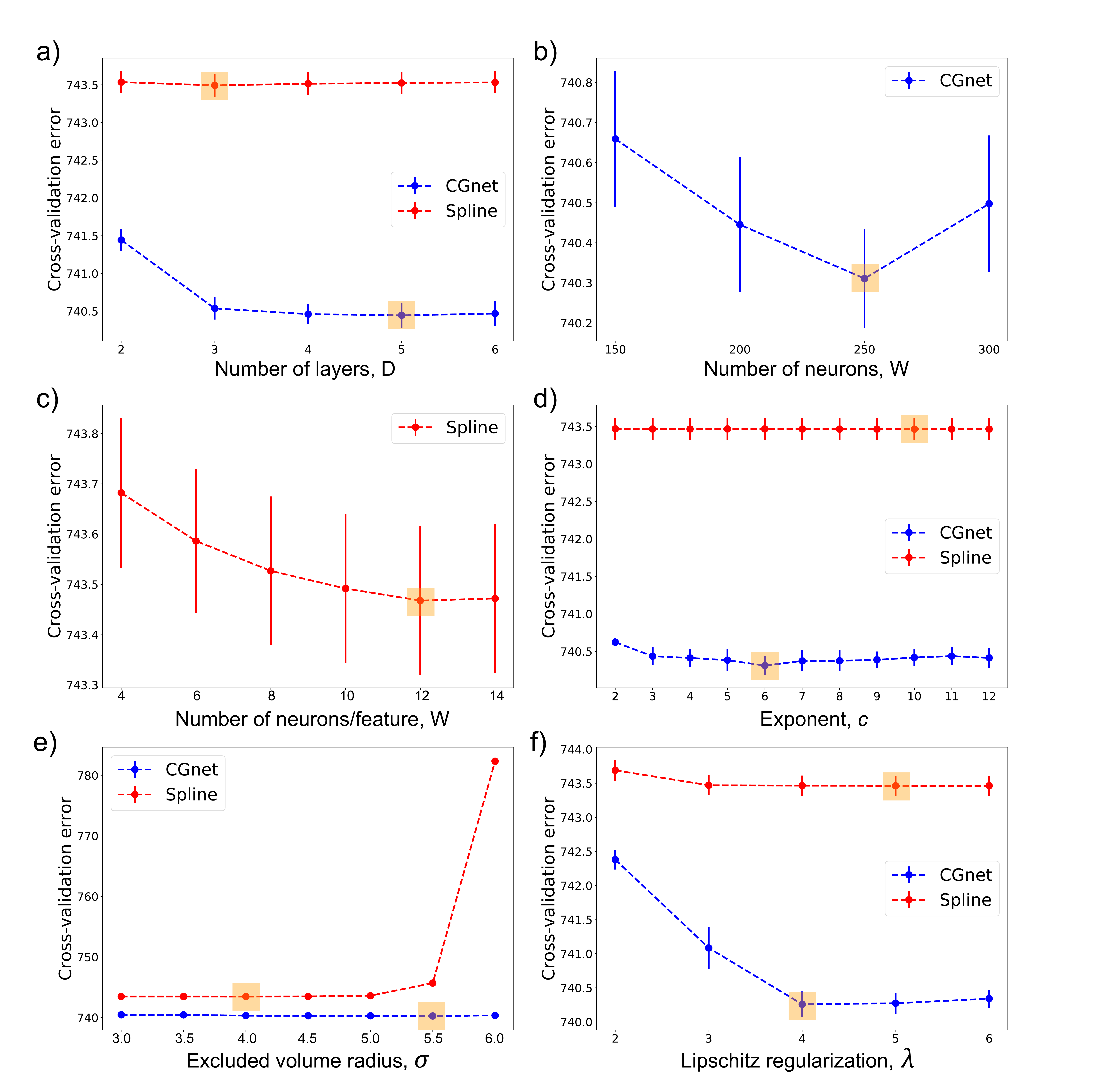}
\par\end{centering}
\caption{\label{fig:Five-stage-cross-validation-Chignolin}
\footnotesize Five-stage cross-validation
of the hyper-parameters for the CG models of Chignolin. 
(a) Selection of the number of layers, D. (b) and (c) Selection of the number of
neurons per layer, W. (d) Selection of the exponent of the excluded volume
term, c. (d) Selection of the effective excluded volume radius,
$\sigma$. (f) Selection of of the Lipschitz regularization strength, $\lambda$.
The optimal
values are indicated by orange squares and are used to generate the
results reported in Fig. \ref{fig:Free-energy-Chignolin}.}
\end{figure}


\end{document}